\documentclass[journal]{IEEEtran}

\usepackage{cite}
\usepackage{amsmath,amssymb,amsfonts}
\usepackage{algorithmic}
\usepackage{graphicx}
\usepackage{textcomp}
\usepackage{xcolor}
\usepackage{stfloats}
\usepackage{mathrsfs}
\usepackage{balance}
\usepackage{amsthm} 
\usepackage[ruled,vlined]{algorithm2e}
\newtheorem{theorem}{Theorem}
\newtheorem{proposition}{Proposition}
\newtheorem{corollary}{Corollary}[theorem] 
\newtheorem{remark}{Remark}

\usepackage{floatrow}
\usepackage{comment}
\usepackage{makecell}
\usepackage{pgfplots}
\pgfplotsset{compat=newest}
\usepackage{acronym}
\usepackage{glossaries}
\usepackage[hidelinks]{hyperref}
\makeglossaries
\newacronym{OTFS}{OTFS}{orthogonal time frequency space}
\newacronym{RCP}{RCP}{reduced cyclic prefix}
\newacronym{ISAC}{ISAC}{integrated sensing and communications}
\newacronym{DD}{DD}{delay-Doppler}
\newacronym{DL}{DL}{deep learning}
\newacronym{SNR}{SNR}{signal-to-noise ratio}
\newacronym{CRLB}{CRLB}{Cramér-Rao lower bound}
\newacronym{NMSE}{NMSE}{normalized mean squared error}
\newacronym{LMMSE}{LMMSE}{linear minimum mean squared error}
\newacronym{V2I}{V2I}{vehicle-to-infrastructure}
\newacronym{V2V}{V2V}{vehicle-to-vehicle}
\newacronym{OFDM}{OFDM}{orthogonal frequency division multiplexing}
\newacronym{PAPR}{PAPR}{peak-to-average power ratio}
\newacronym{ISI}{ISI}{intersymbol interference}
\newacronym{ICI}{ICI}{intercarrier interference}
\newacronym{MF}{MF}{matched filter}
\newacronym{MFs}{MFs}{matched filters}
\newacronym{DDs}{DDs}{delays-Dopplers}
\newacronym{OMP}{OMP}{orthogonal matching pursuit}
\newacronym{SBL}{SBL}{sparse Bayesian learning}
\newacronym{IPI}{IPI}{interpath interference}
\newacronym{DDIPIC}{DDIPIC}{delay-Doppler interpath interference cancellation}
\newacronym{PIPIC}{PIPIC}{progressive interpath interference cancellation}
\newacronym{GLRT}{GLRT}{generalized likelihood ratio test}
\newacronym{MIMO}{MIMO}{multiple input multiple output}
\newacronym{MP}{MP}{message passing}
\newacronym{MRC}{MRC}{maximal ratio combining}
\newacronym{IMFC}{IMFC}{iterative matched filtering and combining}
\newacronym{BER}{BER}{bit error rate}
\newacronym{AWGN}{AWGN}{additive white Gaussian noise}
\newacronym{PSD}{PSD}{power spectral density}
\newacronym{CAF}{CAF}{cross-ambiguity function}
\newacronym{CP}{CP}{cyclic prefix}
\newacronym{SFFT}{SFFT}{symplectic finite Fourier transform}
\newacronym{LOS}{LOS}{line-of-sight}
\newacronym{RTD}{RTD}{round trip delay}
\newacronym{HSR}{HSR}{high-speed railways}
\newacronym{SC}{SC}{stopping criterion}
\newacronym{ISFFT}{ISFFT}{inverse symplectic finite Fourier transform}
\newacronym{DFT}{DFT}{discrete Fourier transform}
\newacronym{I/O}{I/O}{input-output}
\newacronym{DDRE}{DDRE}{delay-Dopppler resource element}
\newacronym{CUT}{CUT}{cell under test}
\newacronym{FNN}{FNN}{feedforward neural network}
\newacronym{ReLU}{ReLU}{rectified linear unit}
\newacronym{PDP}{PDP}{power delay profile}
\newacronym{FIM}{FIM}{Fisher information matrix}
\newacronym{DPS}{DPS}{Doppler power spectrum}
\newacronym{iid}{iid}{independent identically distributed}
\newacronym{MFC}{MFC}{matched filtering and combining}
\newacronym{MSE}{MSE}{mean squared error}
\newacronym{QAM}{QAM}{quadrature amplitude modulation}
\newacronym{RMSE}{RMSE}{root mean squared error}
\newacronym{ML}{ML}{maximum likelihood}
\newacronym{CSI}{CSI}{channel state information}
\newacronym{TM}{TM}{threshold method}
\newacronym{CM-FNN}{CM-FNN}{FNN-based correlation method}
\newacronym{DZT}{DZT}{discrete Zak transform}
\newacronym{IDZT}{IDZT}{inverse discrete Zak transform}
\newacronym{TSE}{TSE}{two step estimator}
\newacronym{LS}{LS}{least squares}
\newacronym{CDDPM}{CDDPM}{constituent delay-Doppler parameter matrix}
\newacronym{LoS}{LoS}{line-of-sight}

\usepackage{etoolbox}
\usepackage{orcidlink}

\allowdisplaybreaks

\begin{document}

\title{Exploiting Structural Sparsity and Delay-Doppler Decoupling for Low-Complexity \\OTFS-ISAC Receivers}

\author{
Mauro Marchese\orcidlink{0009-0008-0265-5840},~\IEEEmembership{Graduate Student Member,~IEEE}, Musa Furkan Keskin\orcidlink{0000-0002-7718-8377},~\IEEEmembership{Member,~IEEE}, \\ Pietro Savazzi\orcidlink{0000-0003-0692-8566},~\IEEEmembership{Senior Member,~IEEE}, and Henk Wymeersch\orcidlink{0000-0002-1298-6159},~\IEEEmembership{Fellow,~IEEE}%
\thanks{M. Marchese is with the Department of Electrical, Computer and Biomedical Engineering, University of Pavia, 27100 Pavia, Italy (e-mail: mauro.marchese01@universitadipavia.it).}
\thanks{Pietro Savazzi is with the Department of Electrical, Computer and Biomedical Engineering, University of Pavia, Pavia, 27100 Italy (e-mail: pietro.savazzi@unipv.it), and with the Consorzio Nazionale Interuniversitario per le Telecomunicazioni - CNIT.}
\thanks{M. F. Keskin and H. Wymeersch are with the Department of Electrical Engineering, Chalmers University of Technology, 412 96 Gothenburg, Sweden.}
\thanks{This work is supported, in part, by Vinnova FFI project 2022-01640 (Beyond 5GPOS), by the Swedish Research Council (VR) through the project 6GPERCEF under Grant 2024-04390, and by the European Union under the Italian National Recovery and Resilience Plan (NRRP) of NextGenerationEU, partnership on “Telecommunications of the Future” (PE00000001 - program "RESTART”).}
}

\maketitle

\begin{abstract}
In this work, the problems of channel estimation, radar sensing, and data detection are addressed for monostatic \gls{ISAC} applications within \gls{OTFS} systems operating with a \gls{RCP}. 
Specifically, the \gls{DD} input-output relationship is formulated in a discrete representation that enables signal-independent disjoint parameter estimation by encapsulating fractional delay and Doppler effects through distinct, structurally sparse matrices. 
This exact algebraic separability is directly exploited to develop a low-complexity parameter estimation framework for the communication channel, which is seamlessly adapted for monostatic radar sensing on backscattered data frames. 
To enhance path detection robustly and safeguard estimation accuracy under low \gls{SNR} regimes where traditional \gls{SC}-based methods fail, a \gls{DL} architecture is integrated to perform model order selection via multi-class classification. 
Furthermore, a path-wise variant of the iterative Landweber method, designated as \gls{IMFC}, is introduced for low-complexity data detection by leveraging the identical structural sparsity unlocked by the decoupled framework. 
Simulation results indicate the proposed estimation scheme achieves lower \gls{NMSE} than conventional channel estimation algorithms and sensing performance close to the \gls{CRLB}.
Finally, the \gls{IMFC} equalizer is shown to deliver \gls{BER} performance comparable to the traditional \gls{LMMSE} benchmark while dramatically reducing the computational load.
\end{abstract}

\begin{IEEEkeywords}
OTFS, integrated sensing and communications (ISAC), parameter estimation, deep learning (DL), reduced-complexity equalization.
\end{IEEEkeywords}

\glsresetall

\section{Introduction}
\IEEEPARstart{H}{igh-mobility} scenarios are becoming increasingly prevalent with the ongoing development of next-generation wireless networks \cite{Giordani2020,Zhang2019}. As such, novel waveforms have been investigated to accommodate these emerging use cases, including \gls{HSR}, \gls{V2I}, and \gls{V2V} communications where speeds up to $500$ km/h are experienced \cite{Hadani2017}. The conventional \gls{OFDM} scheme adopted in 4G/5G systems robust to \gls{ISI} becomes impractical in such conditions, as its performance degrades significantly in high-Doppler channels \cite{Wang2006} due to \gls{ICI}. 
\Gls{OTFS} is a 2D modulation scheme proposed in \cite{Hadani2017,HadaniMonk2018} and is considered a strong candidate waveform for 6G networks due to its resilience to high-Doppler effects in high-mobility scenarios. \gls{OTFS} has been shown \cite{Hadani2017} to deliver superior performance in high-Doppler environments. Furthermore, \gls{OTFS} is well-suited for sensing due to its inherent delay-Doppler channel representation \cite{Mohammed2022,Gaudio2020}. Moreover, \gls{OTFS} exhibits lower \gls{PAPR} compared to \gls{OFDM} \cite{Wei2022papr,serrano2025}. This encourages considering an \gls{ISAC} system where a dual-functional transmitter is used for both communication and sensing. 
While \gls{OTFS} demonstrates significant potential in high-Doppler scenarios, its practical implementation hinges on the development of efficient algorithms for channel estimation and data detection. Channel estimation plays a pivotal role as the receiver's ability to decode data depends on an accurate representation of the channel. Therefore, in order to enhance the feasibility of \gls{OTFS} for next-generation wireless \gls{ISAC} systems, \textit{novel low-complexity algorithms must be developed for  channel estimation, radar sensing, and data detection}. 

\subsection{State of the Art}

\subsubsection{Channel Estimation}
Several algorithms have been proposed in the literature for \gls{OTFS} channel estimation. In \cite{Raviteja2019,Viterbo2022}, a threshold method was proposed, achieving good performance in both integer \gls{DDs} and fractional Doppler scenarios. Channel sparsity is exploited in \cite{Rasheed2022} to design an \gls{OMP} algorithm for \gls{OTFS} channel estimation. \Gls{SBL} algorithms for channel estimation have also been explored in \cite{Zhao2020,Wei2022,Zhang2024sbl,Li2025}. Low-complexity channel estimation schemes, enabling disjoint delay-Doppler estimation in fractional \gls{DDs} scenarios, are proposed in \cite{Khan2021,Khan2023,Yogesh2024}. Specifically, the low-complexity disjoint delay-Doppler estimation schemes in \cite{Khan2021,Khan2023} achieve decoupling by strictly relying on the transmission of an isolated impulse pilot with an appropriately sized guard zone. In such frameworks, the delay-Doppler separability is a signal-dependent property observed at the receiver side, which vanishes when arbitrary data symbols are transmitted as usual in \gls{ISAC} systems. In more detail, in \cite{Khan2021,Khan2023}, an approximate \gls{ML} estimation algorithm, derived under the assumption of large \gls{OTFS} frame size, is proposed. In fact, if this condition holds, multipaths are well separated in the \gls{DD} domain and the columns of the \gls{CDDPM} \cite{Muppaneni2023,marchese2024} become orthogonal \cite{Khan2021,Khan2023}. Under the assumption of good separability of received pilot replicas due to fine delay and Doppler resolutions, the algorithms in \cite{Khan2021,Khan2023} outperform the methods in \cite{Raviteja2019,Rasheed2022} and \gls{SBL} methods. When delay and Doppler resolutions are not fine enough, received pilot replicas spread into adjacent \gls{DD} bins, causing non-negligible \gls{IPI} that limits estimation accuracy. To this end, channel estimation in high-\gls{IPI} scenarios has been addressed in \cite{Muppaneni2023,marchese2024}. In \cite{Muppaneni2023}, a \gls{DDIPIC} algorithm is proposed, employing refinement procedures to cancel \gls{IPI}. This approach outperforms the method from \cite{Khan2021}, at the cost of higher complexity. Conversely, in \cite{marchese2024}, a variant of the \gls{DDIPIC} algorithm, called \gls{PIPIC}, is introduced to reduce complexity and enhance estimation accuracy. It is shown that a single global refinement step is sufficient to achieve high estimation accuracy and suppress \gls{IPI}, surpassing the performance of \gls{DDIPIC}. However, despite their resilience in high-\gls{IPI} scenarios, the algorithms in \cite{Muppaneni2023,marchese2024} perform joint \gls{DD} estimation.

Nevertheless, the methodologies proposed in \cite{Khan2021,Khan2023,Muppaneni2023,marchese2024} are based on a \gls{DD} input/output relation that accounts for fractional channel parameters and is quite complicated since it stems from continuous-time analysis \cite{Gaudio2020,Khan2021,Khan2023}. A more compact expression for the \gls{DD} \gls{I/O} relation in matrix form is provided in \cite{Raviteja2019_2}. This expression is valid for \gls{OTFS} with a single \gls{CP} (referred to in the literature as \gls{RCP}-\gls{OTFS} \cite{Raviteja2019_2,Viterbo2022}) and assumes integer delays and Doppler shifts. A compact \gls{DD} \gls{I/O} relation accounting for fractional \glspl{DD} has been derived in \cite{Wu2023}. However, the relation between the expression proposed in \cite{Wu2023} has not been related to the well-known integer \gls{DD} based expression from \cite{Raviteja2019_2}. Moreover, in \cite{Wu2023}, a joint \gls{DD} estimation algorithm has been developed for fractional \gls{DD} channels.
Finally, the approaches in \cite{marchese2024,Muppaneni2023,Mattu2024,Khan2021,Khan2023,Yogesh2024} use stopping criteria to detect multiple paths. This may lead to poor estimation performance at low \gls{SNR} where small amplitude paths may be missed.

Since \gls{DL} has emerged as a powerful tool for channel estimation, offering reduced computational cost and latency while improving accuracy, several \gls{DL}-based approaches have been proposed in the \gls{OTFS} literature to enhance channel state information extraction \cite{Mattu2022,Gao2024,Guo2023}, improve robustness to noise \cite{Zhang2022_adth,Zhang2022_AN}, refine model-based estimates \cite{Li2022}, and reduce computational complexity \cite{Mattu2024,marchese2025reducedlat}.

\subsubsection{Radar Sensing}
As \gls{OTFS} exhibits good properties for sensing \cite{serrano2025,song2025}, several approaches have been investigated.
In \cite{Muppaneni2023}, the \gls{DDIPIC} algorithm is shown to achieve good sensing performance in an \gls{ISAC} context in which the transmitter estimates the radar parameters of the target receiver. Although in \cite{Muppaneni2023} a single-target scenario has been considered, the \gls{DDIPIC} algorithm can be used for multi-target detection and parameter estimation. Other radar sensing algorithms in single-antenna systems are considered in \cite{raviteja2019Radar,Keskin2021}, where \gls{MF} radar and \gls{GLRT} based algorithms are proposed, respectively. In \cite{Keskin2021}, \gls{ISI} and \gls{ICI} effects are exploited to overcome the ambiguity barrier of conventional sensing algorithms in \gls{OTFS} systems. \gls{OTFS}-based \gls{ISAC} in \gls{MIMO} systems has been considered in \cite{Dehkordi2022,Gaudio2020RadarMIMO,Keskin2024,singh2025}. 
However, such approaches entail joint \gls{DD} estimation of radar target parameters \cite{Muppaneni2023,Keskin2021,Dehkordi2022,Gaudio2020RadarMIMO,Keskin2024}, thereby compounding the computational burden through a multi-dimensional search space.

\subsubsection{Data Detection}
Various approaches have been investigated to reduce complexity and enhance detection performance. Similar to \gls{OFDM}, simple single-tap equalization is considered in \cite{Viterbo2022}, but its performance is acceptable only in quasi-static (very low-mobility) multipath channels. For high-mobility scenarios, the traditional \gls{LMMSE} estimator \cite{Viterbo2022} can be employed to achieve better performance. In \cite{Surabhi2020,Tiwiri2019,Naikoti2021_det,Pfadler2021}, various \gls{LMMSE} detector variants have been proposed to mitigate the high computational complexity caused by the matrix inversion required in \gls{LMMSE}, while maintaining good detection performance. However, biorthogonality \cite{Surabhi2020} and integer channel parameters \cite{Tiwiri2019} are assumed.
A more efficient approach, exploiting sparsity in the \gls{DD} domain, is presented in \cite{Raviteja2018,Viterbo2022}. In this case, the detection problem is formulated as a sparsely connected factor graph, which can be efficiently solved using the \gls{MP} algorithm. 
However, the main advantage of the \gls{MP} algorithm comes from the sparsity of the \gls{DD} channel matrix in integer \gls{DDs} scenarios. On the contrary, when fractional channel parameters are experienced, the \gls{DD} channel matrix becomes denser and complexity increases.

\subsection{Contributions}
In this work, the problem of channel estimation, radar sensing and data detection in \gls{RCP}-\gls{OTFS} systems with fractional \gls{DDs} is investigated.
Given the above description of the existing \gls{OTFS} literature, the following points are addressed in this paper: \textit{(i)} formulation of a compact expression for the \gls{DD} channel matrix accounting for fractional channel parameters and enabling signal-independent separate \gls{DD} estimation, whereas conventional disjoint \gls{DD} estimation algorithms strictly rely on an impulse-pilot-based assumption \cite{Khan2021,Khan2023}; \textit{(ii)} enhancing path detection and channel estimation accuracy at low \gls{SNR} values, where conventional \gls{SC}-based approaches miss small amplitude paths; \textit{(iii)} design of a reduced-complexity equalization algorithm for fractional \gls{DD} case that relies on the intrinsic properties of \gls{DD} parameter matrices by relaxing the assumptions made in \cite{Tiwiri2019,Surabhi2020}. Thus, the contributions of this work are as follows: 
\begin{itemize}  

    \item \textbf{Signal-independent algebraic delay-Doppler separability framework:} Unlike existing disjoint fractional \gls{DD} estimation methods \cite{Khan2021,Khan2023} that rely on continuous-time formulations, asymptotic approximations valid only for large frame sizes, and signal-dependent parameter matrices embedded with a impulse pilot signal, a rigorous, closed-form algebraic decomposition of the \gls{RCP}-\gls{OTFS} channel matrix is developed. It is demonstrated that fractional delay and Doppler effects can be rigorously decoupled into distinct, structurally sparse parameter matrices. This separability is completely independent of the transmitted signal, ensuring that \gls{DD} decoupling and structural sparsity can be exploited also in data-based \gls{ISAC} (i.e., when random data symbols are transmitted). 
    Leveraging this separability, a correlation-based parameter estimation algorithm is proposed for \gls{OTFS}-\gls{ISAC}, which reduces the multi-dimensional search over the \gls{DD} domain into two independent, one-dimensional optimization problems. 

    \item \textbf{Enhanced model order selection at low \gls{SNR} via \gls{DL}}: As conventional channel estimation algorithms rely on a \gls{SC} to detect multiple paths, performance heavily depends on the choice of the convergence tolerance parameter for the \gls{SC}, and at low \gls{SNR}, multiple paths are often missed. To enhance detection capabilities and estimation accuracy at low \gls{SNR}, a \gls{DL} architecture is introduced to estimate the number of propagation paths by formulating the estimation problem as a multi-class classification task.

    \item \textbf{Reduced-complexity Landweber method for channel equalization}: A path-wise version of the Landweber method, exploiting sparsity of DD parameter matrices and \gls{DD} separability, is proposed for data detection to reduce the equalization complexity against the \gls{LMMSE} estimator. This algorithm, called \gls{IMFC}, iteratively combines the outputs of \gls{MFs} parametrized by different delays/Dopplers to estimate data symbols. This approach offers significantly lower complexity than the traditional \gls{LMMSE} estimator while achieving nearly the same \gls{BER} performance.  
\end{itemize}

\textit{Notation}: $\|\mathbf{X}\|_F$, represents the Frobenius norm. $\mathbf{I}_N$ is the identity matrix of order $N$. The operators $\text{vec}(\mathbf{X})$ and $\text{vec}_{M,N}^{-1}(\mathbf{x})$ represent the vectorization of matrix $\mathbf{X}$ and the reshaping of vector $\mathbf{x}$ back into a $M\times N$ matrix, respectively. $\otimes$, $\odot$ and $\times$ represent the Kronecker (tensor), Hadamard (element-wise) and Cartesian products, respectively. Finally, $\rho(\cdot)$ denotes the spectral radius, which is the maximum absolute eigenvalue of the argument, and $[\cdot]_A$ denotes the modulo-$A$ operation.

\begin{figure}[t]
\centering
    \includegraphics[width=\textwidth, trim=80 160 140 160, clip]{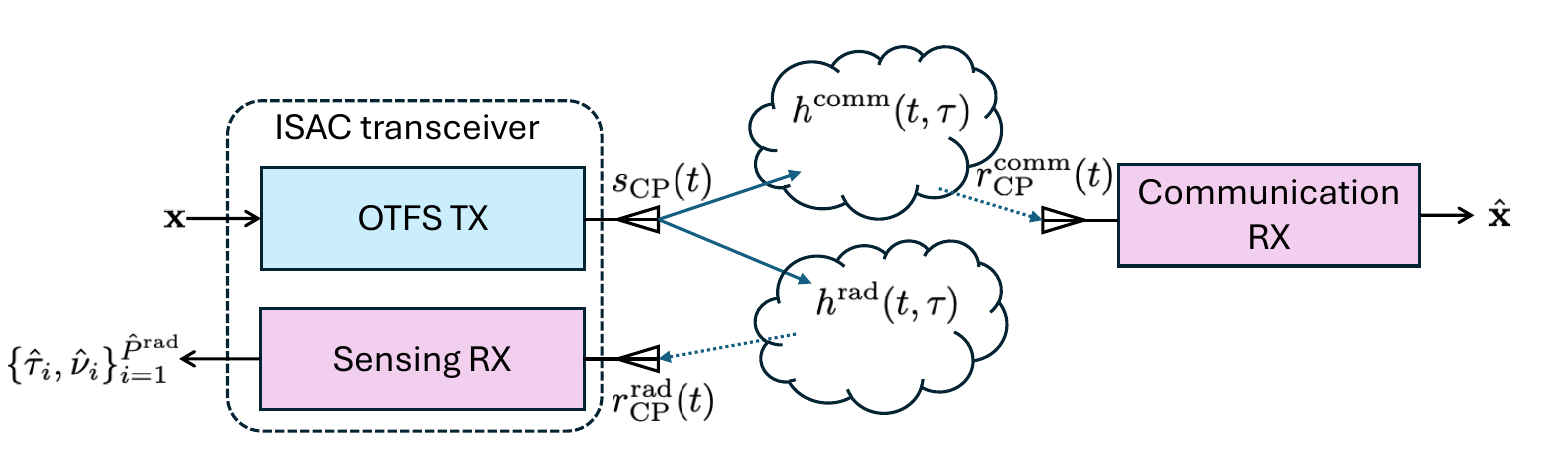}
    \caption{The RCP-OTFS ISAC system including: a dual-functional ISAC transceiver for data communication and radar sensing (OTFS TX and Sensing RX are mounted on the same hardware) and a communication receiver.}\label{fig:ISACsystem}
\end{figure}

\section{RCP-OTFS System Model}\label{sysmodel}
The considered \gls{ISAC} scenario includes a dual-functional OTFS transceiver (made of a single-antenna transmitter for communication and a radar sensing receiver mounted on the same hardware platform) and a single-antenna communication receiver, as shown in Fig.~\ref{fig:ISACsystem}. 
In this scenario, it is assumed that there exists a geometric coherence time during which the channel parameters (channel gains, delays and Doppler shifts) can be considered constant \cite{Viterbo2022}. Typically, these geometric coherence intervals of the \gls{DD} channel are significantly longer than the transmission duration of individual \gls{OTFS} frames \cite{Viterbo2022}. Specifically, while the complex path gains inherently undergo deterministic phase rotations from one frame to the next due to mobility, they can be treated as invariant across consecutive frames modulo this phase shift, which is known a priori and easily compensated for once the Doppler shifts are estimated. Therefore, in order to avoid interference between pilot and data symbols due to fractional delays and Dopplers, the transmitter sends, within this geometric coherence window, a single pilot frame for channel estimation, followed by a sequence of data frames \cite{Muppaneni2023,Mattu2024,Khan2021,marchese2024,marchese2025reducedlat}.
It is worth noting that while conventional embedded pilot-aided strategies are widely used for integer channels, they become highly inefficient in fractional delay-Doppler regimes \cite{Viterbo2022}. Under an embedded setup, mitigating the resulting severe interference would mandate an excessively large guard zone around the pilot impulse, thereby drastically increasing the transmission overhead and undermining the spectral efficiency\cite{Viterbo2022}. Conversely, the considered separate pilot-and-data frame strategy entirely bypasses data-to-pilot interference without incurring prohibitive overhead penalties, taking full advantage of the long geometric coherence time of the \gls{DD} channel.
A numerical assessment that justifies the considered separate pilot and data frames is provided in Section \ref{sec:SimScenario}, where the frame duration is compared with the typical geometric coherence intervals of the \gls{DD} channel.

At the receiver side, the communication receiver processes the pilot frame for estimating the communication channel. Afterwards, it uses the channel estimate to detect incoming data frames. In the meanwhile, the transmit signal is reflected off multiple targets in the environment. The backscattered data frames are used by the sensing receiver to estimate range and velocity of the targets \cite{Muppaneni2023}.

\subsection{Transmit Signal Model}
The \gls{OTFS} system is characterized by $M$ subcarriers and $N$ time slots. The subcarrier spacing is $\Delta f = 1/T$, where $T$ represents the OTFS block duration. Therefore, the signal bandwidth is $M\Delta f$, and the overall frame duration is $NT$. The \gls{OTFS} modulator arranges $MN$ information symbols in the \gls{DD} domain over the two-dimensional grid $I_{DD}=\big\{m\Delta\tau, n\Delta\nu \ | \ 0 \leq m \leq M-1, \ 0 \leq n \leq N-1\big\}$, where $\Delta\tau = 1/(M\Delta f)$ and $\Delta\nu = 1/(NT)$ represent the delay and Doppler resolutions, respectively. Hence, the \gls{DD} matrix $\mathbf{X} \in \mathbb{C}^{M \times N}$ is obtained. The information symbols are selected from an alphabet with $Q$ constellation points.
The $n$-th \gls{OTFS} symbol is given by the Heisenberg transform as
\begin{equation} \label{eq:HeisTransform}
s_n(t)=\sum_{m=0}^{M-1}\big[\mathbf{X}_{\text{tf}}\big]_{m,n}e^{j2\pi m\Delta ft}g_{\text{tx}}(t), \ \ 0<t<T,
\end{equation}
where $g_{\text{tx}}(t)$ is the transmit pulse shaping waveform, and $\mathbf{X}_{\text{tf}} \in \mathbb{C}^{M \times N}$ is the time-frequency samples matrix. The delay-Doppler symbols are converted to time-frequency samples using the \gls{ISFFT} as
$\mathbf{X}_{\text{tf}}=\mathbf{F}_M\mathbf{X}\mathbf{F}_{N}^{\mathsf{H}}$
where $[\mathbf{F}_N]_{p,q}=\frac{1}{\sqrt{N}}e^{-j2\pi\frac{pq}{N}}$ is the $N$-point unitary \gls{DFT} matrix.
The entire time-domain \gls{OTFS} signal is then given by
\begin{equation} \label{eq:OTFSsignal}
s(t)=\sum_{n=0}^{N-1}s_n(t-nT).
\end{equation}
In \gls{RCP}-\gls{OTFS} \cite{Viterbo2022,Raviteja2019_2} systems the overall transmitted signal is preceded by a single cyclic prefix of duration $T_{\text{CP}}$. Hence, the transmitted \gls{RCP}-\gls{OTFS} signal is given as
\begin{align} \label{eq:CPsignal}
s_{\text{CP}}(t)=
    \begin{cases}
    s(t), & 0\leq t\leq NT \\
    s(t+NT), & -T_{\text{CP}}\leq t\leq 0
    \end{cases}.
\end{align}
The duration of the \gls{CP} is chosen such that $T_{\text{CP}}\geq\sigma_{\tau}$, where $\sigma_{\tau}$ is the channel delay spread.
Sampling \eqref{eq:HeisTransform} at $t=qT/M$ with $q=0,1,\dots,M-1$ and stacking samples into a vector, \eqref{eq:HeisTransform} can be expressed in matrix form as  
\begin{equation} \label{eq:DiscreteHeis}
\mathbf{s}_n=\mathbf{G}_{\text{tx}}\mathbf{F}_M^{\mathsf{H}}\big[\mathbf{X}_{\text{tf}}\big]_{:,n} \ ,
\end{equation}
where $\mathbf{G}_{\text{tx}}=\text{diag}\Big\{g_{\text{tx}}(0),g_{\text{tx}}\Big(\frac{T}{M}\Big),...,g_{\text{tx}}\Big(\frac{(M-1)T}{M}\Big)\Big\}$.
The overall transmit samples are then obtained as
\begin{equation} \label{eq:TransmitSamples}
\mathbf{S}=[\mathbf{s}_1 \ \mathbf{s}_2 \ \dots \ \mathbf{s}_N]=\mathbf{G}_{\text{tx}}\mathbf{F}_M^{\mathsf{H}}\mathbf{X}_{\text{tf}}=\mathbf{G}_{\text{tx}}\mathbf{X}\mathbf{F}_N^{\mathsf{H}}.
\end{equation}
By vectorizing \eqref{eq:TransmitSamples}, the transmit samples vector is given by
\begin{equation} \label{eq:TransmitSamplesVector}
\mathbf{s}=\text{vec}(\mathbf{S})=\big(\mathbf{F}_{N}^{\mathsf{H}}\otimes\mathbf{G}_{\text{tx}}\big)\text{vec}(\mathbf{X})=\big(\mathbf{F}_{N}^{\mathsf{H}}\otimes\mathbf{G}_{\text{tx}}\big)\mathbf{x}.
\end{equation}
In case of rectangular pulse shaping $\mathbf{G}_{\text{tx}}=\mathbf{I}_{M}$, thus \eqref{eq:TransmitSamplesVector} reduces to the well-known \gls{IDZT} \cite{Viterbo2022}. Hereafter, \gls{OTFS} with rectangular pulse shaping is considered as it outperforms other pulse shaping waveforms \cite{Raviteja2019_2}. 

\subsection{High-mobility Communication and Radar Channel}
The high-mobility communication and radar channel is made of a certain number of mobile scatterers. Therefore, the delay-time impulse response of the channel with $P$ reflectors can be obtained as\footnote{Here it is assumed that the time-bandwidth product is low such that Doppler-induced time-scaling effects can be neglected \cite{Keskin2024} (i.e., $s(t-\tau(t))\approx s(t-\tau)$, where $\tau(t)=\tau-\nu t/f_c$ is the Doppler-induced time-varying delay). Conversely, frequency-domain shifts caused by the Doppler effect cannot be neglected, leading to the channel model in \eqref{eq:delayTimeChannel}.}
\begin{equation} \label{eq:delayTimeChannel}
h(t,\tau)=\sum_{i=1}^Pg_ie^{j2\pi\nu_i t}\delta(\tau-\tau_i),
\end{equation}
where $g_i$, $\tau_i$ and $\nu_i$ are the complex channel gain, the propagation delay and the Doppler shift of the $i$-th reflected path, respectively. For radar sensing, $\tau_i=\frac{2d_i}{c}$ is the \gls{RTD} where $d_i$ is the target range and $c$ is the speed of light; and $\nu_i=\frac{2v}{c}f_c$ where $v$ is the radial velocity of the target and $f_c$ is the carrier frequency.

\subsection{Received Signal Model}
The received signal at both communication and sensing receivers is expressed as
\begin{equation} 
\begin{split}\label{eq:RxMultipathCP}
r_{\text{CP}}(t)&=\int h(t,\tau)s_{\text{CP}}(t-\tau)\text{d}\tau+n(t)
\\
&=\sum_{i=1}^Pg_is_{\text{CP}}(t-\tau_i)e^{j2\pi\nu_i t}+n(t),
\end{split}
\end{equation}
where $n(t)$ is \gls{AWGN} with monolateral \gls{PSD} $N_0$.

\section{Novel RCP-OTFS ISAC Model Formulation}
\subsection{Time Domain Analysis for RCP Waveforms}
In this section, the expression for the delay-time channel matrix is derived. Let's consider a multicarrier \gls{RCP} waveform with $MN$ symbols per frame. The received signal is obtained according to \eqref{eq:RxMultipathCP}. The received signal without \gls{CP} is given as \cite{Raviteja2019_2,Keskin2024,Keskin2021}
\begin{equation} \label{eq:RxMultipath}
r(t)=\sum_{i=1}^Pg_is([t-\tau_i]_{NT})e^{j2\pi\nu_i t}+n(t).
\end{equation}
By sampling at the symbol rate, the following discrete observations, expressed in matrix form, are obtained
\begin{equation} \label{eq:DelayTimeIO}
\mathbf{r}=\mathbf{Hs}+\mathbf{n},
\end{equation}
where $\mathbf{H}\in\mathbb{C}^{MN\times MN}$ is the delay-time channel matrix and $\mathbf{n}\sim\mathcal{CN}(\mathbf{0},\sigma^2\textbf{I}_{MN})\in\mathbb{C}^{MN\times 1}$ is the \gls{AWGN} vector.

\begin{theorem}[]\label{Th1}
Given the \gls{I/O} relation in \eqref{eq:DelayTimeIO}, the delay-time channel matrix is given by
\begin{equation} \label{eq:DelayTimeChannelMtx}
\mathbf{H}=\sum_{i=1}^Pg_i\mathbf{D}^{K_i}\mathbf{D}^{\tilde{k}_i}\mathbf{F}_{MN}^{\mathsf{H}}(\mathbf{D}^*)^{L_i}(\mathbf{D}^*)^{\tilde{l}_i}\mathbf{F}_{MN},
\end{equation}
where $\mathbf{D}=\textnormal{diag}\{[z^{q}]_{q=0}^{MN-1}\}$ with $z=e^{j\frac{2\pi}{MN}}$, $L_i=\textnormal{round}(\frac{\tau_i}{\Delta\tau})$ denotes the integer delay, $\tilde{l}_i=\frac{\tau_i-L_i\Delta\tau}{\Delta\tau}$ denotes the fractional delay, $K_i=\textnormal{round}(\frac{\nu_i}{\Delta\nu})$ denotes the integer Doppler and $\tilde{k}_i=\frac{\nu_i-K_i\Delta\nu}{\Delta\nu}$ denotes the fractional Doppler.
\end{theorem}

\begin{proof}
See Appendix A.
\end{proof}

Theorem \ref{Th1} generalizes the results in \cite{Raviteja2019_2} by removing the simplifying assumption of integer channel parameters. It can be noted that a similar expression of the delay-time channel matrix accounting for fractional channel parameters has been proposed in \cite{Wu2023}, without investigating \gls{DD} decoupling.

\begin{corollary}\label{Corol1}
If channel parameters assume integer values \big($\tilde{l}_i=0,\tilde{k}_i=0$\big), then the delay-time channel matrix reduces to 
\begin{equation} \label{eq:DelayTimeChannelMtxIntDDs}
\mathbf{H}=\sum_{i=1}^Pg_i\mathbf{D}^{K_i}\mathbf{\Pi}^{L_i},
\end{equation}
where $\Pi=[\mathbf{e}_2 \ \mathbf{e}_3 \ \dots \ \mathbf{e}_{MN} \ \mathbf{e}_1]$ is the forward cyclic shift permutation matrix and $\mathbf{e}_i$ is a one-hot vector defined as $[\mathbf{e}_i]_q=\delta_{iq}$.
\end{corollary}

\begin{proof}
See Appendix B.
\end{proof}
From Corollary \ref{Corol1}, it is evident that the derived expression for the delay-time channel matrix in \eqref{eq:DelayTimeChannelMtx} serves as a generalization of \eqref{eq:DelayTimeChannelMtxIntDDs} presented in \cite{Raviteja2019_2}. Furthermore, it can be observed that the effects of propagation delays and Doppler shifts are encapsulated through the same diagonal matrix $\mathbf{D}$.

\subsection{Novel Formulation of the Delay-Doppler ISAC Model}
In this section, a novel formulation of the \gls{DD} expression for the \gls{ISAC} channel, enabling disjoint \gls{DD} estimation, is derived based on the result presented in Theorem \ref{Th1}. 

Given the received samples vector $\mathbf{r}$ from \eqref{eq:DelayTimeIO}, the information symbol vector is obtained by inverting \eqref{eq:TransmitSamplesVector} as \cite{Raviteja2019_2,Viterbo2022}
\begin{equation} \label{eq:RxDDsymbolsVector}
\mathbf{y}=\big(\mathbf{F}_{N}\otimes\mathbf{I}_{M}\big)\mathbf{r}.
\end{equation}
As shown in \cite{Raviteja2019_2,Viterbo2022}, combining \eqref{eq:TransmitSamplesVector}, \eqref{eq:DelayTimeIO} and \eqref{eq:RxDDsymbolsVector} the \gls{I/O} relation in the \gls{DD} domain can be written as
\begin{equation} \label{eq:IODelayDoppler}
\mathbf{y}=\mathbf{H}_{\text{DD}}\mathbf{x}+\mathbf{w},
\end{equation}
where $\mathbf{w}=\big(\mathbf{F}_{N}\otimes\mathbf{I}_{M}\big)\mathbf{n}$ is the \gls{AWGN} noise vector in the \gls{DD} domain and 
\begin{equation} \label{eq:DDChannelMtx}
\mathbf{H}_{\text{DD}}=\big(\mathbf{F}_{N}\otimes\mathbf{I}_{M}\big)\mathbf{H}\big(\mathbf{F}_{N}^{\mathsf{H}}\otimes\mathbf{I}_{M}\big),
\end{equation}
is the \gls{DD} channel matrix.
It can be noticed that $\mathbf{w}$ is zero mean with covariance matrix
\begin{equation}\begin{split}\label{eq:CovarianceNoiseDD}
\mathbf{C}_{\text{w}}&=\mathbb{E}\big[\mathbf{w}\mathbf{w}^{\mathsf{H}}\big]=\big(\mathbf{F}_{N}\otimes\mathbf{I}_{M}\big)\mathbb{E}\big[\mathbf{n}\mathbf{n}^{\mathsf{H}}\big]\big(\mathbf{F}_{N}^{\mathsf{H}}\otimes\mathbf{I}_{M}\big)
\\
&=\sigma^2\big(\mathbf{F}_{N}\mathbf{F}_{N}^{\mathsf{H}}\otimes\mathbf{I}_{M}^2\big)=\sigma^2\big(\mathbf{I}_{N}\otimes\mathbf{I}_{M}\big)=\sigma^2\mathbf{I}_{MN}.
\end{split}
\end{equation}
Replacing \eqref{eq:DelayTimeChannelMtx} in \eqref{eq:DDChannelMtx}, the \gls{DD} channel matrix becomes
\begin{equation}\label{eq:DDChannelMtxNovel}
\mathbf{H}_{\text{DD}}=
\sum_{i=1}^Pg_i\big(\mathbf{F}_{N}\otimes\mathbf{I}_{M}\big)\mathbf{D}^{k_i}\mathbf{F}_{MN}^{\mathsf{H}}(\mathbf{D}^*)^{l_i}\mathbf{F}_{MN}\big(\mathbf{F}_{N}^{\mathsf{H}}\otimes\mathbf{I}_{M}\big),
\end{equation}
where $l_i={\tau_i}/{\Delta\tau}=L_i+\tilde{l}_i $ and $k_i={\nu_i}/{\Delta\nu}=K_i+\tilde{k}_i$ are the normalized \gls{DDs}.
Thus, by defining
\begin{equation}\label{eq:TiMtx}
\mathbf{T}_i
=\big(\mathbf{F}_{N}\otimes\mathbf{I}_{M}\big)\mathbf{D}^{k_i}\mathbf{F}_{MN}^{\mathsf{H}}(\mathbf{D}^*)^{l_i}\mathbf{F}_{MN}\big(\mathbf{F}_{N}^{\mathsf{H}}\otimes\mathbf{I}_{M}\big),
\end{equation}
the \gls{DD} channel matrix is given as
\begin{equation}\label{eq:DDChannelMtxNovelsumgiTi}
\mathbf{H}_{\text{DD}}=\sum_{i=1}^Pg_i\mathbf{T}_i.
\end{equation}
Moreover, it is possible to define the following matrix
\begin{equation}\label{eq:QMtx}
\mathbf{Q}(a)=\big(\mathbf{F}_{N}\otimes\mathbf{I}_{M}\big)\mathbf{D}^a\mathbf{F}_{MN}^{\mathsf{H}}\big(\mathbf{F}_{N}\otimes\mathbf{I}_{M}\big)\in\mathbb{C}^{MN\times MN},
\end{equation}
such that
\begin{equation}\label{eq:TiQQ}
\mathbf{T}_i=\mathbf{Q}(k_i)\mathbf{Q}^*(l_i).
\end{equation}

\begin{remark}\label{rm3}
The effects of the propagation delay and the Doppler shift are separable since it can be observed that the matrix $\mathbf{T}_i$ in \eqref{eq:TiQQ} is the product of a Doppler term $\mathbf{Q}(k_i)$ and a delay term $\mathbf{Q}^*(l_i)$. This notable property of the formulation in \eqref{eq:TiQQ} enables the separate estimation of \gls{DD} parameters. As a result, disjoint estimation can be performed to reduce computational complexity, avoiding a joint bidimensional search over the \gls{DD} domain. Moreover, unlike previous literature where the delay and Doppler terms are decoupled within a measurement matrix \cite{Khan2021,Khan2023} (i.e., the \gls{CDDPM} matrix \cite{Muppaneni2023}) that embeds the pilot signal itself, the algebraic decomposition in \eqref{eq:TiQQ} operates directly on the channel matrix $\mathbf{H}_{\text{DD}}$. Consequently, this matrix factorization holds true irrespective of the nature of the input vector $\mathbf{x}$. This fundamental generalization ensures that separability can be leveraged also when data are transmitted, thereby unlocking disjoint estimation in data-based \gls{ISAC} applications.
\end{remark}
The disjoint parameter estimation is further discussed in Section \ref{ChEst}.

In conclusion, the following unified \gls{ISAC} model for \gls{RCP}-\gls{OTFS} systems is obtained
\begin{equation}\label{eq:ISACModel}
\mathbf{y}=\sum_{i=1}^Pg_i\mathbf{Q}(k_i)\mathbf{Q}^*(l_i)\mathbf{x}+\mathbf{w}.
\end{equation}
This model exhibits some interesting properties highlighted in the following proposition.
\begin{proposition}\label{lem:Qexpression}
The expression for a generic element of $\mathbf{Q}(a)$ is given by
\begin{equation}\begin{split}\label{eq:Qelem_lemma}
\big[\mathbf{Q}(a)\big]_{p,q} &= \frac{1}{N\sqrt{MN}} \sum_{k_1=0}^{N-1} e^{-j\frac{2\pi \lfloor p/M \rfloor k_1}{N}} \\  \sum_{k_2=0}^{N-1} & e^{-j\frac{2\pi k_2 \lfloor q/M \rfloor}{N}}e^{j\frac{2\pi (k_1 M + \text{mod}(p,M))(a+k_2 M+\text{mod}(q,M))}{MN}}.
\end{split}
\end{equation}
Moreover, if $M$ is an integer multiple of $N$ (typical in \gls{OTFS} system design), the number of non-zero elements in each row is $M$ and the positions of that elements are $q=M\text{mod}(p,N)+m$ where $m=0,1,\dots,M-1$.
\end{proposition}
\begin{proof}
See Appendix C.
\end{proof}

\begin{remark}
Proposition \ref{lem:Qexpression} reveals the sparse structure of the $\mathbf{Q}(a)$ matrix in the proposed \gls{DD} channel model in \eqref{eq:ISACModel}. In particular, it can be noted that the computation of each element of $\mathbf{Q}(a)$ has complexity $\mathcal{O}(N^2)$. Thus, the overall complexity for the computation of $\mathbf{Q}(a)$ is $\mathcal{O}(M^2N^3)$. This property is then exploited in Section \ref{ChEst} for designing a low-complexity channel parameter estimation algorithm. It is important to mention that this sparsity stems only from the definition of the $\mathbf{Q}(a)$ matrix in \eqref{eq:QMtx} and does not intertwine with that of the \gls{DD} channel matrix in \eqref{eq:DDChannelMtx}, which arises from the nature of wireless channel in the \gls{DD} domain \cite{Hadani2017,Viterbo2022}. Furthermore, the sparsity of the \gls{DD} channel matrix depends on the quantization resolution along the delay and Doppler axes \cite{Chong2025}; consequently, it cannot be exploited in the same manner as the structural sparsity of the $\mathbf{Q}(a)$ matrix.

\end{remark}
 The structure of the $\mathbf{Q}(a)$ matrix is shown in Figure \ref{fig:Qmatrix}, where it can be noted that, with a fractional value of $a$, there are only $M$ nonzero elements per row.

\begin{figure}[t]
    \centering
    \resizebox{0.99\columnwidth}{!}{
        \input{PaperResults/tikz_files/matriceQ.tikz}}
    \caption{The magnitude of the elements of $\mathbf{Q}(a)$ for $M=16$, $N=4$ and $a=4.4$}
    \label{fig:Qmatrix}
\end{figure}

\section{Channel Estimation and Radar Sensing}\label{ChEst}
In this section, a novel channel estimation and radar sensing algorithm is developed by leveraging the unitary and the sparsity properties of $\mathbf{Q}(a)$. The goal of channel estimation is to determine the channel parameters ($P, g_i, \tau_i, \nu_i$) by processing the received pilot signal. In this section, the pilot signal model is introduced and the proposed channel estimation algorithm is presented conditional on knowledge of $P$. Subsequently, the estimation of $P$ is discussed and the proposed technique is extended for radar sensing at the \gls{ISAC} transceiver.

\subsection{Pilot Signal Model for Channel Estimation at Communication Receiver}\label{sec:PilotModel}
The single-antenna transmitter sends a pilot symbol for channel estimation in the \gls{DD} domain. The \gls{DD} pilot frame is given by \cite{Muppaneni2023,Mattu2024,Khan2021,marchese2024,marchese2025reducedlat}
\begin{align} \label{eq:PilotModel}
[\mathbf{X}_p]_{m,n}=
    \begin{cases}
    \sqrt{E_p} & m=m_p,~n=n_p \\
    0 & \textnormal{otherwise}
    \end{cases},
\end{align}
where ($m_p,n_p$) is the \gls{DDRE} in which the pilot is sent and $E_p$ is the energy of the transmitted pilot signal. After transmission, assuming that the channel gains are normalized such that $\sum_{i=1}^P |g_i|^2 = 1$, the received pilot signal power is equal to the transmitted pilot signal power \cite{Khan2021,marchese2024}. The pilot signal power is defined as the pilot energy divided by the frame duration, resulting in $E_p / NT$. On the other hand, the noise power is calculated as the product of the noise spectral density $N_0$ and the signal bandwidth $M\Delta f$. Consequently, the pilot signal-to-noise ratio is expressed as:
\begin{equation}\label{eq:PilotSNR}
\text{SNR}_p=\frac{E_p}{MNN_0}.
\end{equation}
Equivalently, $\text{SNR}_p=E_p/\text{tr}(\mathbf{C}_{\text{w}})=E_p/MN\sigma^2$. Thus, the noise variance is $\sigma^2=N_0$.

\subsection{Proposed Correlation-based Channel Estimation Method Conditional on Knowledge of P}\label{Sec:ProposedCMmethod}
As previously discussed, a low-complexity channel estimation algorithm can be developed by leveraging the sparsity of the $\mathbf{Q}(a)$ matrix through the correlation of the pilot via a bank of \gls{MFs}. Furthermore, disjoint \gls{DD} estimation can be performed due to the separability property highlighted in Remark \ref{rm3}. This allows the two-dimensional search over \gls{DD} domain to be reduced to two one-dimensional search along delay and Doppler. For the time being, we assume the number of propagation paths, $P$, is known. 

The channel parameters of multiple paths can be estimated as follows:
\begin{enumerate}
\item \textit{DD initialization}:
The integer part of the channel parameters can be easily obtained by identifying the maximum pilot energy in the received frame. The search space is constrained to $\mathcal{S}=[0,\ L_{\max}]\times[-K_{\max},\ K_{\max}]$. Thus, the following estimates are obtained
\begin{equation} \label{eq:DDinitialization}
\hat{L}_i,\hat{K}_i=\arg\max_{(L,K)\in\mathcal{S}}\Big|[\mathbf{Y}^{(i-1)}]_{m_p+L,n_p+K}\Big|^2,
\end{equation}
where $\mathbf{Y}^{(i-1)}=\text{vec}_{M,N}^{-1}(\mathbf{y}^{(i-1)})$. In the first iteration, $\mathbf{y}^{(0)}=\mathbf{y}$. 

\item \textit{Doppler refinement}:
Once the integer part is estimated, the delay is fixed at $\hat{L}_i$, while the Doppler is refined by estimating the fractional part. Specifically, the observation is passed through a filterbank of \gls{DDs} parameterized by different Dopplers within the error range. Then, a \gls{MF} parameterized by the fixed delay is applied. In this manner, the \gls{DD} pilot is reconstructed as $\mathbf{Q}^\top(\hat{L}_i)\mathbf{Q}^{\mathsf{H}}(\hat{K}_i + \tilde{k})\mathbf{y}^{(i-1)}$ for different values of $\tilde{k}\in[-\frac{1}{2},\frac{1}{2}]$. The estimate for the fractional Doppler is obtained by maximizing the magnitude of the correlation between the pilot $\mathbf{x}_p$ and the reconstructed pilot. Specifically,
\begin{equation} \label{eq:DopplerRefinement}
\hat{\tilde{k}}_i=\arg\max_{\tilde{k}\in[-\frac{1}{2},\frac{1}{2}]}\Big|\mathbf{x}_p^{\mathsf{H}}\mathbf{Q}^\top(\hat{L}_i)\mathbf{Q}^{\mathsf{H}}(\hat{K}_i+\tilde{k})\mathbf{y}^{(i-1)}\Big|.
\end{equation}
This refinement can be performed hierarchically to further improve accuracy by quantizing the error range in each iteration. By following this procedure for $L_h$ iterations, the estimation error can be progressively reduced. Specifically, during the $h$-th iteration, the new estimate is obtained as
\begin{equation}\begin{split} \label{eq:HierarchicalDopplerRefinement}
&\hat{\tilde{k}}_i^{(h)}=\hat{\tilde{k}}_i^{(h-1)}+
\\
&\Delta k^{(h)}\arg\max_{\kappa}\Big|\mathbf{x}_p^{\mathsf{H}}\mathbf{Q}^\top\big(\hat{L}_i\big)\mathbf{Q}^{\mathsf{H}}\Big(\hat{K}_i+\kappa\Delta k^{(h)}\Big)\mathbf{y}^{(i-1)}\Big|,
\end{split}
\end{equation}
where the number of grid points is $2N_k+1$ and $\kappa\in[-N_k,N_k]$. During the $h$-th iteration the resolution is $\Delta k^{(h)}=1/(2N_k)^{h}$. Finally, the estimate $\hat{k}_i=\hat{K}_i+\hat{\tilde{k}}_i$ is obtained.

\item \textit{Matched filtering}:
Once the Doppler is refined, a Doppler-compensated version of the observation is obtained by passing $\mathbf{y}^{(i-1)}$ through a \gls{MF} parametrized by the refined Doppler. Thus, 
\begin{equation} \label{eq:ObservationMatchedFiltering}
\mathbf{y}_d^{(i-1)}=\mathbf{Q}^{\mathsf{H}}(\hat{k}_i)\mathbf{y}^{(i-1)}.
\end{equation}

\item \textit{Delay refinement}:
The delay is refined by estimating the fractional part. Specifically, the Doppler-compensated observation $\mathbf{y}_d^{(i-1)}$ is passed through a filterbank of \gls{MFs}, each parametrized by different delays spanning the error range. The reconstructed \gls{DD} pilot is then given by $\mathbf{Q}^\top(\hat{L}_i + \tilde{l}) \mathbf{y}_d^{(i-1)}$ for different values of $\tilde{l}\in[-\frac{1}{2},\frac{1}{2}]$. The estimate for the fractional delay is obtained by maximizing the magnitude of the correlation between the pilot $\mathbf{x}_p$ and the reconstructed pilot. Thus,
\begin{equation} \label{eq:DelayRefinement}
\hat{\tilde{l}}_i=\arg\max_{\tilde{l}\in(-\frac{1}{2},\frac{1}{2})}\Big|\mathbf{x}_p^{\mathsf{H}}\mathbf{Q}^\top(\hat{L}_i+\tilde{l})\mathbf{y}_d^{(i-1)}\Big|.
\end{equation}
Similarly to the Doppler case, the delay can be refined hierarchically for $L_h$ times. During the $h$-th iteration the new estimate is obtained as
\begin{equation}\begin{split} \label{eq:HierarchicalDelayRefinement}
&\hat{\tilde{l}}_i^{(h)}=\hat{\tilde{l}}_i^{(h-1)}+
\\
&\Delta l^{(h)}\arg\max_{\ell}\Big|\mathbf{x}_p^{\mathsf{H}}\mathbf{Q}^\top(\hat{L}_i+\ell\Delta l^{(h)})\mathbf{y}_d^{(i-1)}\Big|,
\end{split}\end{equation}
where the number of grid points is $2N_l+1$ and $\ell\in[-N_l,N_l]$ if $\hat{\tilde{l}}_i^{(h-1)}\neq 0$, otherwise the number of grid points is $N_l+1$ and $\ell\in[0,N_l]$. During the $h$-th iteration the resolution is $\Delta l^{(h)}=1/(2N_l)^{h}$. Finally, the estimate $\hat{l}_i=\hat{L}_i+\hat{\tilde{l}}_i$ is obtained.

\item \textit{Gain estimate}:
Once the \gls{DD} parameters of the path are obtained $\big(\hat{\mathbf{T}}_i=\mathbf{Q}(\hat{k}_i)\mathbf{Q}^*(\hat{l}_i)\big)$, a low-complexity estimate of the channel gain can be derived by solving \eqref{eq:ISACModel} for $g_i$ in the noiseless single-path case. Thus, the channel gain estimate is given as the orthogonal projection of $\mathbf{y}^{(i-1)}$ onto $\hat{\mathbf{T}}_i\mathbf{x}_p$. Thus,
\begin{equation} \label{eq:GainEstimCM}
\hat{g}_i=\frac{(\hat{\mathbf{T}}_i\mathbf{x}_p)^{\mathsf{H}}\mathbf{y}^{(i-1)}}{\|\mathbf{x}_p\|^2}=\frac{(\hat{\mathbf{T}}_i\mathbf{x}_p)^{\mathsf{H}}\mathbf{y}^{(i-1)}}{E_p}.
\end{equation}

\item \textit{\gls{IPI} cancellation}:
In the multipath scenario, the channel parameters of the different paths are estimated by iterating through the previous steps $P$ times. At the end of each iteration, a residual \gls{DD} pilot vector is obtained as
\begin{equation} \label{eq:ResidueComputation}
\mathbf{y}^{(i)}=\mathbf{y}^{(i-1)}-\hat{g}_i\hat{\mathbf{T}}_i\mathbf{x}_p,
\end{equation}
and it is used as the new observation for estimating parameters of the next path \cite{Khan2021,Khan2023,marchese2024,Yogesh2024}. As explained in \cite{marchese2024,Yogesh2024}, this procedure allows for \gls{IPI} cancellation.
\end{enumerate}

The pseudocode of the proposed low-complexity disjoint \gls{DD} correlation-based channel estimation method is provided in Algorithm \ref{alg:CM}.

\begin{remark}
The proposed channel parameter estimation algorithm leverages the sparsity of $\mathbf{Q}(a)$ to reduce computational complexity, as characterized in Proposition \ref{lem:Qexpression}. This sparse structure arises when $M$ is an integer multiple of $N$. Conversely, if this condition is not met, the $\mathbf{Q}(a)$ matrix becomes denser, and the sparsity property can no longer be exploited to reduce computational complexity. Nevertheless, since the unitary property of $\mathbf{Q}(a)$ remains intact, the proposed algorithm remains applicable, albeit at the cost of increased complexity. 
\end{remark}

\begin{algorithm}[t]
\caption{Proposed Correlation-based Method for Channel Estimation}\label{alg:CM}
\KwIn{$\mathbf{y},\ P,\ L_h,\ L_{\max},\ K_{\max}$}
$\mathbf{y}^{(0)}=\mathbf{y}$\\
\For{$i = 1$ \KwTo $P$}{
  $\mathbf{Y}^{(i-1)} = \mathrm{vec}_{M,N}^{-1}\!\left(\mathbf{y}^{(i-1)}\right)$\\
  $\hat{L}_i,\,\hat{K}_i = \arg\max\limits_{(L,K)\in\mathcal{S}}\left|[\mathbf{Y}^{(i-1)}]_{m_p+L,\,n_p+K}\right|^2$\\
  $\hat{\tilde{k}}_i^{(0)}=0$\\
  \For{$h=1$ \KwTo $L_h$}{
    $\Delta k^{(h)}=1/(2N_k)^h$\\
    $\kappa^*=\arg\max\limits_{\kappa}\!\left|\mathbf{x}_p^{\mathsf{H}}\mathbf{Q}^{\top}(\hat{L}_i)\mathbf{Q}^{\mathsf{H}}\!\left(\hat{K}_i+\kappa\Delta k^{(h)}\right)\!\mathbf{y}^{(i-1)}\right|$\\
    $\hat{\tilde{k}}_i^{(h)}=\hat{\tilde{k}}_i^{(h-1)}+\Delta k^{(h)}\kappa^*$\\
  }
  $\hat{k}_i = \hat{K}_i+\hat{\tilde{k}}_i^{(L_h)}$\\
  $\mathbf{y}_d^{(i-1)} = \mathbf{Q}^{\mathsf{H}}(\hat{k}_i)\,\mathbf{y}^{(i-1)}$\\
  $\hat{\tilde{l}}_i^{(0)}=0$\\
  \For{$h=1$ \KwTo $L_h$}{
    $\Delta l^{(h)}=1/(2N_l)^h$ \\
    $\ell^*=\arg\max\limits_{\ell}\!\left|\mathbf{x}_p^{\mathsf{H}}\mathbf{Q}^{\top}\!\left(\hat{L}_i+\ell\Delta l^{(h)}\right)\!\mathbf{y}_d^{(i-1)}\right|$\\
    $\hat{\tilde{l}}_i^{(h)}=\hat{\tilde{l}}_i^{(h-1)}+\Delta l^{(h)}\ell^*$\\
  }
  $\hat{l}_i = \hat{L}_i+\hat{\tilde{l}}_i^{(L_h)}$\\
  $\hat{\mathbf{T}}_i = \mathbf{Q}(\hat{k}_i)\,\mathbf{Q}^*(\hat{l}_i)$\\
  $\hat{g}_i = \frac{\bigl(\hat{\mathbf{T}}_i\mathbf{x}_p\bigr)^{\mathsf{H}}\!\mathbf{y}^{(i-1)}}{E_p}$\\
  $\mathbf{y}^{(i)} = \mathbf{y}^{(i-1)}-\hat{g}_i\,\hat{\mathbf{T}}_i\mathbf{x}_p$\\
}
\textbf{Output:}~$\hat{\mathbf{H}}_{\mathrm{DD}}=\sum_{i=1}^{P}\hat{g}_i\hat{\mathbf{T}}_i$\\
\end{algorithm}

\subsection{Estimation of P: DL-Aided Model Order Selection}\label{DLpathdet}
In practical \gls{ISAC} scenarios, the number of reflectors $P$ is unknown. Iterative channel estimation algorithms often rely on a \gls{SC} \cite{marchese2024,Muppaneni2023,Mattu2024,Khan2021,Khan2023,Yogesh2024} to estimate $P$ and detect multiple paths. A conventional approach relies on defining a threshold based on the energy of the residual vector in \eqref{eq:ResidueComputation}. Specifically, when the energy of $\mathbf{y}^{(i)}$ falls below a designated tolerance value, the iterative process terminates, and the estimated number of propagation paths, $\hat{P}$, is determined by the total number of iterations performed. However, this approach raises the following concerns:

\begin{enumerate}
\item \textit{Threshold-dependent accuracy}:  
The performance of an iterative channel estimation algorithm that uses a \gls{SC} to estimate $P$ heavily depends on the choice of the convergence tolerance parameter. To minimize estimation error, an exhaustive search for the optimal threshold is often necessary \cite{Khan2021}.

\item \textit{IPI-induced estimation/detection trade-off}:  
When a \gls{SC}-based approach is used, a lower threshold may reduce channel estimation error, but it can also lead to an overestimation of $P$. Conversely, a threshold optimized to correctly identify the number of paths may not be ideal for minimizing the overall estimation error. This discrepancy arises because \gls{IPI} limits the accuracy of the estimated paths. In more detail, \gls{IPI} leads to residual pilot peaks that may be interpreted as small amplitude paths, leading to an overestimation of $P$ if the threshold is not properly set. Therefore, a refinement procedure should be carried out to discard false alarms \cite{marchese2024}, increasing complexity.

\item \textit{Poor detection capabilities at low SNR}:  
At low signal-to-noise ratios, small-amplitude paths may be missed if their amplitudes become comparable to the noise level. This occurs because a few high-amplitude paths may be sufficient to meet the stopping criterion, thereby masking the presence of weaker paths.
\end{enumerate}

To this end, the following \gls{DL}-aided scheme is proposed to estimate $P$. The estimation of $P$ is formulated as a single-label multi-class classification problem: a multipath channel belongs to the class $c \in \mathcal{C}$ if it is characterized by $P^{(c)}$ propagation paths, where $\mathcal{C}$ denotes the set of considered model orders. A \gls{FNN} is trained to solve this classification task. Specifically, the element-wise magnitude of the received pilot vector, denoted as $|\mathbf{y}|=\mathbf{y}\odot\mathbf{y^*} \in \mathbb{R}^{MN \times 1}$, is provided as input to the network. This choice is physically motivated by the fact that the key information regarding the number of propagation paths is inherently embedded in the power profile of the received pilot signal, where distinct multipath components manifest as energy peaks in the delay-Doppler domain, whereas the phase information is highly volatile and less indicative of the model order. 

To formalize the processing pipeline of the \gls{FNN}, let $L$ denote the total number of layers. The real-valued feature vector $|\mathbf{y}|$ is fed into the first hidden layer, yielding the activation vector $\mathbf{a}^{(1)} \in \mathbb{R}^{N_1 \times 1}$
\begin{equation}
\mathbf{a}^{(1)} = \sigma_{\text{ReLU}} \left( \mathbf{W}^{(1)} |\mathbf{y}| + \mathbf{b}^{(1)} \right),
\end{equation}
where $\mathbf{W}^{(1)} \in \mathbb{R}^{N_1 \times MN}$ and $\mathbf{b}^{(1)} \in \mathbb{R}^{N_1 \times 1}$ represent the weight matrix and the bias vector of the first layer, respectively, while $\sigma_{\text{ReLU}}(\cdot) = \max(0, \cdot)$ denotes the element-wise \gls{ReLU} activation function. For any subsequent hidden layer $l \in \{2, \dots, L-1\}$, the operational model transitions recursively as
\begin{equation}
\mathbf{a}^{(l)} = \sigma_{\text{ReLU}} \left( \mathbf{W}^{(l)} \mathbf{a}^{(l-1)} + \mathbf{b}^{(l)} \right),
\end{equation}
where $\mathbf{W}^{(l)} \in \mathbb{R}^{N_l \times N_{l-1}}$ and $\mathbf{b}^{(l)} \in \mathbb{R}^{N_l \times 1}$ are the weights and biases corresponding to the $l$-th layer. 

To map the output of the final hidden layer into a mathematically sound posterior probability distribution over the $C = |\mathcal{C}|$ classes, the output layer utilizes the Softmax activation function. The resulting network output vector $\mathbf{p} \in \mathbb{R}^C$ is mathematically defined as
\begin{equation}
\mathbf{p} =  \sigma_{\text{soft}} \left( \mathbf{W}^{(L)} \mathbf{a}^{(L-1)} + \mathbf{b}^{(L)} \right),
\end{equation}
where $\mathbf{W}^{(L)} \in \mathbb{R}^{C \times N_{L-1}}$ and $\mathbf{b}^{(L)} \in \mathbb{R}^{C \times 1}$ denote the parameters of the final layer, and the $c$-th element of $\mathbf{p}$ is computed via the Softmax operator $\sigma_{\text{soft}}(\cdot)$ as follows
\begin{equation}
[\mathbf{p}]_c = \frac{e^{[\mathbf{W}^{(L)} \mathbf{a}^{(L-1)} + \mathbf{b}^{(L)}]_c}}{\sum_{k=1}^{C} e^{[\mathbf{W}^{(L)} \mathbf{a}^{(L-1)} + \mathbf{b}^{(L)}]_k}}.
\end{equation}
The term $[\mathbf{p}]_c$ represents the estimated probability that the channel contains a number of paths corresponding to the $c$-th class. After training, the final estimate of $P$ is obtained during inference via the following selection rule
\begin{equation} \label{eq:Pestimation}
\hat{P}=\arg\max_{c\in\mathcal{C}}[\mathbf{p}]_c .
\end{equation}

\begin{remark}
It is worth noting that if the actual number of propagation paths in the channel exceeds the maximum cardinality predefined during the design phase, the proposed \gls{FNN} will inevitably saturate, leading to a degradation in estimation accuracy. However, this behavior does not represent a specific drawback of the \gls{DL}-based approach; rather, it constitutes an intrinsic limitation of parametric channel estimation frameworks in general. Even in conventional \gls{SC}-based methods, a maximum allowable number of paths must be strictly enforced \cite{marchese2024,Muppaneni2023,Mattu2024,Khan2021,Khan2023,Yogesh2024}. This hard limit is practically necessary not only to bound computational complexity but also to prevent algorithmic divergence, such as infinite iterative loops triggered by noise-induced residual peaks. Consequently, the saturation under an unmodeled number of paths is a structural constraint shared by both the \gls{SC}-based baselines and the proposed scheme, reflecting the fundamental trade-offs inherent to parametric channel modeling.    
\end{remark}

\subsection{Multi-Target Parameter Estimation at Sensing Receiver}
\label{sec:multitarget}
In the considered \gls{ISAC} scenario, the transmitted signal is backscattered by a given number of sensing targets in the environment. The radar parameters (range and velocity) of the sensing targets can be estimated at the sensing receiver by processing the received signal in \eqref{eq:ISACModel} using the correlation method.  
The target range and velocity can be obtained by estimating the \gls{RTD} and Doppler frequency associated with the target. In particular, since the co-located sensing receiver has access to all transmit data symbols in \eqref{eq:ISACModel}, the integer \gls{DDs} can be estimated as
\begin{equation} \label{eq:DDestimRadarSensing}
\hat{L}_i,\hat{K}_i=\arg\max_{(L,K)\in\mathcal{S}}\Big|\mathbf{x}^{\mathsf{H}}\mathbf{Q}^\top(L)\mathbf{Q}^{\mathsf{H}}(K)\mathbf{y}^{(i-1)}\Big|,
\end{equation}
by maximizing the magnitude of the correlation between the transmitted data frame $\mathbf{x}$ and the estimated data frame $\mathbf{Q}^\top(L)\mathbf{Q}^{\mathsf{H}}(K)\mathbf{y}^{(i-1)}$ obtained as output of a MF. Afterwards, the fractional delay and Doppler of the target are estimated and refined separately using the disjoint hierarchical approach in \eqref{eq:HierarchicalDopplerRefinement} and \eqref{eq:HierarchicalDelayRefinement}. Multi-target detection is achieved by using \eqref{eq:Pestimation} and iterating this procedure as described in Section \ref{Sec:ProposedCMmethod}. Finally, ranges and velocities of the targets can be obtained using
\begin{equation}
\hat{d}_i=\frac{\hat{l}_i\Delta\tau}{2}c,\ \hat{v}_i=\frac{\hat{k}_i\Delta\nu}{2f_c}c.
\end{equation}

\subsection{Complexity}
Table~\ref{Table:complex} presents a complexity comparison between the proposed channel estimation method, referred to as \gls{CM-FNN}, the high-\gls{IPI}-resilient \gls{PIPIC} approach proposed in \cite{marchese2024}, and the \gls{TSE} algorithm proposed in \cite{Khan2021,Khan2023}. The complexity of the proposed method is primarily driven by the hierarchical refinement of channel parameters. Specifically, the complexity of the proposed method increases with the sum of $N_k$ and $N_l$ due to the separate \gls{DD} estimation, similarly to the \gls{TSE} method \cite{Khan2021,Khan2023}. On the other hand, the complexity of channel estimation algorithms that perform joint \gls{DD} estimation (e.g., PIPIC) \cite{marchese2024,Mattu2024,Muppaneni2023,Khan2021} grows with the product of the number of delay and Doppler grid points, as they rely on a 2D search.  
Moreover, regarding model order selection, the proposed \gls{FNN}-based approach requires higher complexity than conventional \gls{SC}-based approaches in order to provide a substantial performance gain at low \gls{SNR}.

\subsection{Cram\'{e}r-Rao Lower Bound for Radar Sensing}
To evaluate the fundamental limits of the monostatic radar parameter estimation under arbitrary data transmission, the expected \gls{CRLB} is derived. Without loss of generality, a single-target scenario is considered. The vector of radar kinematic parameters is defined by the $\boldsymbol{\theta} = [\tau, \nu]^T$. Following the normalization assumptions maintained throughout the paper, the complex path gain is normalized to unity ($g = 1$), removing it from the active parameter search space.

The discrete \gls{DD} input-output relationship for the target echo in \eqref{eq:ISACModel} is expressed as 
\begin{equation}
    \mathbf{y} = \boldsymbol{\mu}(l,k) + \mathbf{w},
\end{equation}
where $\boldsymbol{\mu}(l,k) =\mathbf{T}\mathbf{x}= \mathbf{Q}(k)\mathbf{Q}^*(l) \mathbf{x}$ is the noiseless received signal vector.

Since the communication data vector $\mathbf{x}$ is random and varies across frames, the \gls{FIM} $\mathbf{J} \in \mathbb{R}^{2 \times 2}$ is formalized by taking the expectation with respect to the signaling statistics of $\mathbf{x}$
\begin{equation}
    [\mathbf{J}]_{p,q} = \frac{2}{\sigma^2} \mathbb{E}_{\mathbf{x}} \left[ \Re \left\{ \left( \frac{\partial \boldsymbol{\mu}(l,k)}{\partial \theta_p} \right)^H \frac{\partial \boldsymbol{\mu}(l,k)}{\partial \theta_q} \right\} \right].
\end{equation}
Under the assumption of \gls{iid} information symbols, the data covariance matrix evaluates to $\mathbf{C}_{\text{x}}=\mathbb{E}\big[\mathbf{x}\mathbf{x}^{\mathsf{H}}\big]=E_s\mathbf{I}_{MN}$, where $E_s$ is the average energy per symbol. Utilizing the linear expectation property $\mathbb{E}[\mathbf{x}^H \mathbf{A} \mathbf{x}] = \text{tr}(\mathbf{A} \mathbb{E}[\mathbf{x}\mathbf{x}^H])$, the elements of the average $2 \times 2$ kinematic \gls{FIM} can be analytically rewritten as
\begin{equation}
    [\mathbf{J}]_{p,q} = \frac{2 E_s}{\sigma^2} \Re \left\{ \text{tr} \left( \left(\frac{\partial \mathbf{T}}{\partial \theta_p}\right)^H \frac{\partial \mathbf{T}}{\partial \theta_q} \right) \right\}.
\end{equation}

The analytical dependency on the continuous delay and Doppler metrics is embedded within the core diagonal matrix $\mathbf{D}$, as shown by \eqref{eq:QMtx}. Let $ \tilde{\mathbf{F}}= \mathbf{F}_N \otimes \mathbf{I}_M$ and define the phase derivative operator as $\mathbf{D}' = \text{diag}\left(\left[ j\frac{2\pi q}{MN} z^q \right]_{q=0}^{MN-1}\right)$. The partial derivatives of the structural sparse parameter matrices $\mathbf{Q}(k)$ and $\mathbf{Q}^*(l)$ with respect to their normalized arguments are explicitly given by
\begin{equation}
    \frac{\partial \mathbf{Q}(k)}{\partial k} = \tilde{\mathbf{F}} \mathbf{D}' \mathbf{D}^{k - 1} \mathbf{F}_{MN}^H \tilde{\mathbf{F}},
\end{equation}
\begin{equation}
    \frac{\partial \mathbf{Q}^*(l)}{\partial l} = \tilde{\mathbf{F}} (\mathbf{D}')^* (\mathbf{D}^*)^{l - 1} \mathbf{F}_{MN}^H \tilde{\mathbf{F}}.
\end{equation}

Applying the chain rule with respect to the continuous physical parameters $\tau$ and $\nu$, the exact algebraic partial derivatives for the channel transfer matrix $\mathbf{T}$ are obtained
\begin{equation}
    \frac{\partial \mathbf{T}}{\partial \tau} = \frac{1}{\Delta \tau} \mathbf{Q}(k) \left( \frac{\partial \mathbf{Q}^*(l)}{\partial l} \right),
\end{equation}
\begin{equation}
    \frac{\partial \mathbf{T}}{\partial \nu} = \frac{1}{\Delta \nu} \left( \frac{\partial \mathbf{Q}(k)}{\partial k} \right) \mathbf{Q}^*(l).
\end{equation}

Upon constructing the $2 \times 2$ kinematic \gls{FIM}, the fundamental lower bounds for the target range $\hat{d}$ and radial velocity $\hat{v}$ are directly extracted from the diagonal elements of the inverted matrix $\mathbf{J}^{-1}$ via the physical mapping relations
\begin{equation}
    \text{CRLB}(d) = \frac{c^2}{4} [\mathbf{J}^{-1}]_{1,1},
\end{equation}
\begin{equation}
    \text{CRLB}(v) = \frac{c^2}{4 f_c^2} [\mathbf{J}^{-1}]_{2,2}.
\end{equation}

\begin{table*}[t]
\centering
\caption{Complexity comparison for channel estimation.}
\small 
\begin{tabular}{|c|c|c|c|} 
 \hline
 & \textbf{Proposed CM-FNN}  & \textbf{TSE \cite{Khan2021,Khan2023}} & \textbf{PIPIC \cite{marchese2024}}  \\ 
 \hline
Estimation of $P$ & \makecell{FNN-based \\ $\mathcal{O}\big((MN)^2\big)$} & \makecell{SC-based \\  $\mathcal{O}(MN)$} & \makecell{SC-based \\  $\mathcal{O}(MN)$} \\
 \hline 
 Estimation of $L_i,K_i$ & \makecell{joint \\ $\mathcal{O}\big((L_{\max}+1)(2K_{\max}+1)\big)$} & \makecell{joint \\ $\mathcal{O}\big((L_{\max}+1)(2K_{\max}+1)\big)$} & \makecell{joint \\ $\mathcal{O}\big((L_{\max}+1)(2K_{\max}+1)\big)$} \\
 \hline 
 Estimation of $\tilde{l}_i,\tilde{k}_i$ & \makecell{disjoint \\ $\mathcal{O}\big(L_hN_k\big)+\mathcal{O}\big(L_hN_l\big)$} & \makecell{disjoint \\ $\mathcal{O}(N_k)+\mathcal{O}(N_l)$} & \makecell{joint \\ $\mathcal{O}\big(L_hN_lN_k\big)$} \\
 \hline 
 Estimation of $g_i$ & $\mathcal{O}(MN)$  & $\mathcal{O}(MN)$ & $\mathcal{O}(PMN)+\mathcal{O}(P^3)$ \\
 \hline 
 Global refinement  & no & no & yes\\
 \hline 
\end{tabular}\label{Table:complex}
\end{table*}

\section{Data Detection at Communication Receiver}
In the \gls{ISAC} scenario under consideration, the single-antenna communication receiver processes the pilot frame to obtain the estimate $\hat{\textbf{H}}_{\text{DD}}$. It then uses this estimate to equalize incoming data frames within the channel coherence time. The transmitter is assumed to send \gls{iid} information symbols. Therefore, $\mathbf{C}_{\text{x}}=\mathbb{E}\big[\mathbf{x}\mathbf{x}^{\mathsf{H}}\big]=E_s\mathbf{I}_{MN}$, where $E_s$ is the average energy per symbol. Since each symbol encodes $\log_2(Q)$ bits, the average energy per bit is given by $E_b=E_s/\log_2(Q)$. The data signal-to-noise ratio is defined as the ratio between the average signal power and the average noise power
\begin{equation}\label{eq:dataSNR}
\text{SNR}_d=\frac{E_s/NT}{M\Delta fN_0}=\frac{E_s}{N_0}.
\end{equation}  
Equivalently, $\text{SNR}_d=\text{tr}(\mathbf{C}_{\text{x}})/\text{tr}(\mathbf{C}_{\text{w}})=E_s/\sigma^2$, leading to a noise variance of $\sigma^2=N_0$.  

In this section, a low-complexity version of the Landweber method \cite{CharlesByrne_2004,Almahdawi2022} that exploits the sparsity of $\mathbf{Q}(a)$ is presented for \gls{OTFS} detection. In particular, a path-wise version of the Landweber method, named \gls{IMFC}, is developed to reduce complexity. The section is structured as follows: first, the \gls{LMMSE} equalization method is reviewed and considered as the benchmark; then, Landweber method is presented and the proposed \gls{IMFC} algorithm is introduced and analyzed.

\subsection{LMMSE Detector}
The \gls{LMMSE} equalizer has been investigated in different works \cite{Tiwiri2019,Surabhi2020,Pfadler2021} and different variants have been proposed to reduce complexity while maintaining satisfactory detection performance. The traditional \gls{LMMSE} equalizer for \eqref{eq:IODelayDoppler} is given by \cite{Viterbo2022,Wei2021}
\begin{equation} \label{eq:LMMSEcompl}
\hat{\mathbf{x}}=\mathbf{C}_{\text{x}}\mathbf{H}_{\text{DD}}^{\mathsf{H}}\big(\mathbf{H}_{\text{DD}}\mathbf{C}_{\text{x}}\mathbf{H}_{\text{DD}}^{\mathsf{H}}+\mathbf{C}_{\text{w}}\big)^{-1}\mathbf{y},
\end{equation}
where $\hat{\mathbf{x}}$ denotes the estimated information symbols vector and the \gls{DD} channel matrix is given as \eqref{eq:DDChannelMtxNovelsumgiTi}. 
Using \eqref{eq:CovarianceNoiseDD}, \eqref{eq:dataSNR} and the symbol's independence assumption, the LMMSE estimate reduces to
\begin{equation} \label{eq:LMMSE}
\hat{\mathbf{x}}=\mathbf{H}_{\text{DD}}^{\mathsf{H}}\Big(\mathbf{H}_{\text{DD}}\mathbf{H}_{\text{DD}}^{\mathsf{H}}+\text{SNR}_d^{-1}\mathbf{I}_{MN}\Big)^{-1}\mathbf{y}.
\end{equation}

\subsection{Proposed IMFC Detector based on Landweber Method}\label{sec:PropIMFC}
The Landweber method is a well-known iterative method for solving ill-posed linear systems of equations \cite{CharlesByrne_2004,Almahdawi2022}. In particular, given the noisy system in \eqref{eq:IODelayDoppler} and considering that $\mathbf{H}_{\text{DD}}$ is typically not invertible, an estimate of $\mathbf{x}$ can be iteratively refined using the following Landweber update equations
\begin{equation} \label{eq:IMFCupdateEq}
\hat{\mathbf{x}}^{(n)}=\hat{\mathbf{x}}^{(n-1)}+\alpha\mathbf{H}_{\text{DD}}^{\mathsf{H}}\boldsymbol{\mathcal{E}}^{(n-1)},
\end{equation}
\begin{equation} \label{eq:IMFCresidue}
\boldsymbol{\mathcal{E}}^{(n)}=\mathbf{y}-\mathbf{H}_{\text{DD}}\hat{\mathbf{x}}^{(n)},
\end{equation}
where $\alpha$ is the step size and $\boldsymbol{\mathcal{E}}^{(n)}$ is the residue vector. The estimation error at time $n$ is given as
\begin{equation} \label{eq:IMFCerr}
\mathbf{e}^{(n)}=\hat{\mathbf{x}}^{(n)}-\mathbf{x}.
\end{equation}
By replacing \eqref{eq:IMFCresidue} in \eqref{eq:IMFCupdateEq}, the estimate at the $n$-th iteration becomes
\begin{equation} \label{eq:IMFCestimate_niter}
\hat{\mathbf{x}}^{(n)}=\alpha\mathbf{H}_{\text{DD}}^{\mathsf{H}}\mathbf{y}+\big(\mathbf{I}_{MN}-\alpha\mathbf{H}_{\text{DD}}^{\mathsf{H}}\mathbf{H}_{\text{DD}}\big)\hat{\mathbf{x}}^{(n-1)}.
\end{equation}
By applying \eqref{eq:IMFCestimate_niter} recursively and assuming that $\hat{\mathbf{x}}^{(n)}=\mathbf{0}$, the estimate can be written as
\begin{equation} \label{eq:IMFCestimate_nafterrecursion}
\hat{\mathbf{x}}^{(n)}=\sum_{i=0}^{n-1}\big(\mathbf{I}_{MN}-\alpha\mathbf{H}_{\text{DD}}^{\mathsf{H}}\mathbf{H}_{\text{DD}}\big)^i\alpha\mathbf{H}_{\text{DD}}^{\mathsf{H}}\mathbf{y}.
\end{equation}
Thus, it can be noted that, if, as $n\rightarrow\infty$
\begin{equation}\label{eq:convcond}
\rho\big(\mathbf{I}_{MN}-\alpha\mathbf{H}_{\text{DD}}^{\mathsf{H}}\mathbf{H}_{\text{DD}}\big)<1,
\end{equation}
where it is reminded that $\rho\big(\mathbf{I}_{MN}-\alpha\mathbf{H}_{\text{DD}}^{\mathsf{H}}\mathbf{H}_{\text{DD}}\big)$ denotes the spectral radius of $\mathbf{I}_{MN}-\alpha\mathbf{H}_{\text{DD}}^{\mathsf{H}}\mathbf{H}_{\text{DD}}$, the algorithm converges \cite{Almahdawi2022}. 
Equivalently, from \eqref{eq:convcond}, it can be noted that the learning rate parameter ($\alpha$) should satisfy 
\begin{equation}
\alpha<\frac{2}{\rho(\mathbf{H}_{\text{DD}}^{\mathsf{H}}\mathbf{H}_{\text{DD}})},
\end{equation}
to guarantee algorithmic convergence.
Moreover, the geometric series in \eqref{eq:IMFCestimate_nafterrecursion} converges to 
\begin{equation}
\begin{split}
\sum_{i=0}^{+\infty}\big(\mathbf{I}_{MN}-\alpha\mathbf{H}_{\text{DD}}^{\mathsf{H}}\mathbf{H}_{\text{DD}}\big)^i=&
(\mathbf{I}_{MN}-(\mathbf{I}_{MN}-\alpha\mathbf{H}_{\text{DD}}^{\mathsf{H}}\mathbf{H}_{\text{DD}}))^{-1}
\\
=(\alpha\mathbf{H}_{\text{DD}}^{\mathsf{H}}\mathbf{H}_{\text{DD}})^{-1}=&\frac{(\mathbf{H}_{\text{DD}}^{\mathsf{H}}\mathbf{H}_{\text{DD}})^{-1}}{\alpha}.
\end{split}
\end{equation}
Thus,
\begin{equation}\label{eq:IMFCestimate_ninfty}
\hat{\mathbf{x}}^{(\infty)}=\lim_{n\rightarrow\infty}\hat{\mathbf{x}}^{(n)}=(\mathbf{H}_{\text{DD}}^{\mathsf{H}}\mathbf{H}_{\text{DD}})^{-1}\mathbf{H}_{\text{DD}}^{\mathsf{H}}\mathbf{y}
\end{equation}
Therefore, the iterative Landweber method converges to the \gls{LS} solution \cite{CharlesByrne_2004}.
Moreover, replacing \eqref{eq:IMFCerr} in \eqref{eq:IMFCestimate_ninfty}, the output error becomes
\begin{equation} \label{eq:IMFCerrUpdateEqResidue}
\mathbf{e}^{(\infty)}=\lim_{n\rightarrow\infty}{\mathbf{e}}^{(n)}=(\mathbf{H}_{\text{DD}}^{\mathsf{H}}\mathbf{H}_{\text{DD}})^{-1}\mathbf{H}_{\text{DD}}^{\mathsf{H}}\mathbf{w}.
\end{equation}
Hence, denoting $\mathbf{H}_{\text{DD}}^\dagger=(\mathbf{H}_{\text{DD}}^{\mathsf{H}}\mathbf{H}_{\text{DD}})^{-1}\mathbf{H}_{\text{DD}}^{\mathsf{H}}$ the Moore-Penrose pseudoinverse of the channel matrix $\mathbf{H}_{\text{DD}}$, the \gls{MSE} after convergence is 
\begin{equation}\begin{split} \label{eq:IMFCmseComputation}
\mathbb{E}\big[\|\mathbf{e}^{(\infty)}\|^2\big] &=
\mathbb{E}\big[\|\mathbf{H}_{\text{DD}}^\dagger\mathbf{w}\|^2\big] = \mathbb{E}\Big[\text{tr}\Big(\mathbf{H}_{\text{DD}}^\dagger\mathbf{w}\mathbf{w}^{\mathsf{H}}(\mathbf{H}_{\text{DD}}^\dagger)^{\mathsf{H}}\Big)\Big]
\\
=&\text{tr}\Big(\mathbf{H}_{\text{DD}}^\dagger\mathbb{E}\big[\mathbf{w}\mathbf{w}^{\mathsf{H}}\big](\mathbf{H}_{\text{DD}}^\dagger)^{\mathsf{H}}\Big)
=\sigma^2\text{tr}\Big(\mathbf{H}_{\text{DD}}^\dagger(\mathbf{H}_{\text{DD}}^\dagger)^{\mathsf{H}}\Big) \\
=&\sigma^2\text{tr}\Big((\mathbf{H}_{\text{DD}}^{\mathsf{H}}\mathbf{H}_{\text{DD}})^{-1}\Big).
\end{split}
\end{equation}
From \eqref{eq:ISACModel}, it is possible to decompose the Landweber iterative update equations path-wise as follows
\begin{equation} \label{eq:IMFCupdateEq_pathwise}
\hat{\mathbf{x}}^{(n)}=\hat{\mathbf{x}}^{(n-1)}+\alpha\sum_{i=1}^Pg_i^*\mathbf{Q}^\top(l_i)\mathbf{Q}^{\mathsf{H}}(k_i)\boldsymbol{\mathcal{E}}^{(n-1)},
\end{equation}
\begin{equation} \label{eq:IMFCresidue_pathwise}
\boldsymbol{\mathcal{E}}^{(n)}=\mathbf{y}-\sum_{i=1}^Pg_i\mathbf{Q}(k_i)\mathbf{Q}^*(l_i)\hat{\mathbf{x}}^{(n)}.
\end{equation}
Therefore, the Landweber method can be decomposed in $P$ separate branches in which the residue vector passes through a \gls{MF} $\mathbf{Q}^{\mathsf{H}}(k_i)$ tuned at the Doppler affecting the $i$-th path and then it passes through a \gls{MF} $\mathbf{Q}^\top(l_i)$ tuned at the delay affecting the $i$-th path. In this way, $P$ incremental updates are obtained and combined using \gls{MRC} to maximize the \gls{SNR}. This path-wise interpretation of the Landweber method is called \gls{IMFC}. The advantage of this approach is that, from Proposition \ref{lem:Qexpression}, each row of $\mathbf{Q}(a)$ contains only $M$ nonzero elements. Therefore, the complexity of the matrix by vector multiplication requires a complexity of $\mathcal{O}(M^2N)$ rather than $\mathcal{O}\big((MN)^2\big)$ as in the Landweber method.

The proposed IMFC equalizer is summarized in Algorithm \ref{alg:IMFC}. For practicality, the step size $\alpha$ is obtained using a time-decaying model as $\alpha^{(n+1)}=\frac{\alpha_0}{1+\beta n}$, where $\alpha_0$ is the initial step size and $\beta$ is the decaying factor. In this way, convergence is ensured by progressively reducing the step size. Moreover, the procedure is iterated for a finite number of steps until $\|\boldsymbol{\mathcal{E}}^{(n)}\|<\epsilon$ or $n=n_{\max}$.

\begin{algorithm}[t]
\caption{Proposed IMFC Detector}\label{alg:IMFC}
\KwIn{$\mathbf{y},\ \epsilon,\ n_{\max},\ \alpha_0,\ \beta$}
$\hat{\mathbf{x}}^{(0)}=\mathbf{0},\ \boldsymbol{\mathcal{E}}^{(0)}=\mathbf{y}, \ \alpha=\alpha_0, \ n=0$\\
\While{$n \leq n_{\max} \ \ \text{AND} \  \ \|\boldsymbol{\mathcal{E}}^{(n)}\| \geq \epsilon$}{
$n = n + 1$\\
\For{$i = 1$ \KwTo $P$}{
  $\boldsymbol{\mathcal{E}}_{D,i}^{(n-1)}=\mathbf{Q}^{\mathsf{H}}(k_i)\boldsymbol{\mathcal{E}}^{(n-1)}$ \\
  $\boldsymbol{\mathcal{E}}_{d,i}^{(n-1)}=\mathbf{Q}^\top(l_i)\boldsymbol{\mathcal{E}}_{D,i}^{(n-1)}$
}
$\hat{\mathbf{x}}^{(n)}=\hat{\mathbf{x}}^{(n-1)}+\alpha\sum_{i=1}^Pg_i^*\boldsymbol{\mathcal{E}}_{d,i}^{(n-1)}$\\
\For{$i = 1$ \KwTo $P$}{
  $\hat{\mathbf{x}}_{d,i}^{(n)}=\mathbf{Q}^*(l_i)\hat{\mathbf{x}}^{(n)}$ \\
  $\hat{\mathbf{x}}_{D,i}^{(n)}=\mathbf{Q}(k_i)\hat{\mathbf{x}}_{d,i}^{(n)}$
}
$\boldsymbol{\mathcal{E}}^{(n)}=\mathbf{y}-\sum_{i=1}^Pg_i\hat{\mathbf{x}}_{D,i}^{(n)}$\\
$\alpha=\frac{\alpha_0}{1+\beta n}$\\}
\textbf{Output:}~$\hat{\mathbf{x}}$\\
\end{algorithm}

\subsection{Complexity: LMMSE vs IMFC}\label{sec:complexityIMFC}
The complexity of the traditional \gls{LMMSE} equalizer is in the order of $\mathcal{O}\big((MN)^3\big)$ due to matrix inversion. On the other hand, the complexity of the Landweber method is $\mathcal{O}\big(n(MN)^2\big)$ and increases linearly with the number of iterations. Instead, the path-wise \gls{IMFC} equalizer exploits the sparsity of $\mathbf{Q}(a)$ as described in Section \ref{sec:PropIMFC}. Therefore, the overall complexity is $\mathcal{O}(nPM^2N)$. It can be concluded that, if $n\ll MN^2/P$, the proposed \gls{IMFC} has much lower complexity than the traditional \gls{LMMSE} equalizer. Moreover, in typical practical settings, it is observed that $P<N$. Thus, the path-wise \gls{IMFC} detector effectively lowers the complexity of the Landweber method. Furthermore, compared to standard iterative solvers that generally incur a computational complexity of at least $\mathcal{O}\big((MN)^2\big)$ per iteration \cite{Albreem2021}, the proposed \gls{IMFC} achieves a reduced computational load, incurring a complexity of $\mathcal{O}(PM^2N)$ per iteration. This reduction in complexity stems from the path-wise operations which directly exploit the structural sparsity of the $\mathbf{Q}(a)$ matrix, as opposed to iterative solvers that rely on the entire channel matrix $\mathbf{H}_{\text{DD}}$, which in the general case of fractional channel parameters is not sparse.

\section{Simulation Results}
In this section, numerical results are reported to validate the proposed estimation/detection methodologies. the transmitter works at carrier frequency $f_c=5$ GHz. The subcarrier spacing is $\Delta f=15$ kHz, hence the time slot duration is $T=66.67 \ \mu$s.

\subsection{Simulation Scenario}\label{sec:SimScenario}
A channel with $P=4$ Rayleigh faded paths is considered. The \gls{PDP} is $[0,\ 2.4,\ 5,\ 7]\mu$s with relative powers $[0, -1, -5,  -7]$ dB . The Doppler shifts of all paths are generated assuming Jakes' \gls{DPS} using $\nu_i=\nu_{\max}cos(\theta_i)$ where $\theta_i\sim\mathcal{U}[0,2\pi]$. The maximum user equipment (UE) speed is $v_{\max}=500$ km/h leading to a maximum Doppler spread of $\nu_{\max}=\frac{v_{\max}}{c}f_c=2.3164$ kHz.
Unless otherwise stated, the \gls{OTFS} frame is characterized by $M=64$ subcarriers and $N=16$ time slots. Grid parameters for estimation in Algorithm~\ref{alg:CM} are set to $N_l,N_k=7$\footnote{The parameters $N_k, N_l$, and $L_h$ are interconnected: a smaller grid requires more hierarchical levels to reach the same resolution, while a larger $L_h$ increases the recursive overhead. The minimum density of the initial grid is constrained by the Hessian of the cost function, as it must be sufficient to capture the main lobe of the objective function to ensure convergence of subsequent refinement steps. Here, $N_k, N_l$ are chosen according to this reasoning and to optimize estimation accuracy while lowering complexity.} and convergence parameters of Algorithm~\ref{alg:IMFC} are set to $\alpha_0=1$ and $\beta=0.05$. Within this scenario, the system operates as described in Section \ref{sysmodel}\footnote{In our simulations, the frame duration is set to $NT\approx1$~ms. Given that the geometric coherence time of the channel parameters (i.e., path gains, delays, and Doppler shifts) is typically on the order of tens of milliseconds \cite{gong2024,Viterbo2022}, these parameters remain effectively stationary across multiple consecutive frames. This physical property justifies the adopted transmission strategy where a single pilot frame is followed by multiple data frames.}.

\subsection{Network Architecture and Training}\label{sec:FNNtraining}
This section details the training process for the proposed \gls{FNN} model order selection scheme. The training dataset $\mathcal{D}$ is a collection of $S$ independent samples. Each sample consists of a pair containing the received pilot observation vector and its corresponding ground-truth label, represented as a one-hot vector
\begin{equation}
    \mathcal{D} = \Big\{(|\mathbf{y}^{(s)}|, \mathbf{p}^{(s)})\Big\}_{s=1}^{S},
\end{equation}
where $|\mathbf{y}^{(s)}| \in \mathbb{C}^{MN \times 1}$ is the $s$-th received pilot frame and $\mathbf{p}^{(s)} \in \{0,1\}^{C}$ is the label vector such that $[\mathbf{p}^{(s)}]_c = 1$ if the $s$-th channel realization contains $P^{(c)}$ paths.
The \gls{FNN} parameters are optimized by minimizing the categorical cross-entropy loss function over the training set
\begin{equation}
    \mathcal{L}(\boldsymbol{\theta}) = -\frac{1}{S} \sum_{s=1}^{S} \sum_{c=1}^{C} [\mathbf{p}^{(s)}]_c \ln([\hat{\mathbf{p}}^{(s)}]_c),
\end{equation}
where $\boldsymbol{\theta}$ denotes the set of network weights and biases, and $\hat{\mathbf{p}}^{(s)} = f_{\text{FNN}}(|\mathbf{y}^{(s)}|; \boldsymbol{\theta})$ is the probability vector predicted by the output layer.

The \gls{DL} network for model order selection is assumed to have two hidden layers. The input dimension is $MN$, as $|\mathbf{y}|$ is provided as input, while the output dimension is $C$, as the vector $\mathbf{p}$ is provided as output. The number of neurons per hidden layer is $\big\{\frac{MN}{4}, \frac{MN}{8}\big\}$, and each hidden layer employs a \gls{ReLU} activation function. The output layer utilizes a softmax activation function to compute probabilities, following the classification problem formulation.  
For the simulations, the proposed \gls{DL} architecture is trained on a dataset comprising $3\times 6000$ samples, collected from three different datasets (each containing $6000$ samples) generated at different $\text{SNR}_p$ levels ($5$ dB, $10$ dB, and $15$ dB, respectively). Each dataset is created under the assumption of channels with $P\sim\mathcal{U}[2,5]$, a uniform \gls{PDP}, and Jakes' \gls{DPS}.
The training process runs for $2000$ epochs with a mini-batch size of $1000$. The learning rate is initially set to $0.001$ and is reduced by a factor of $0.9$ every $50$ epochs.  
This simple architecture for the proposed \gls{DL}-aided scheme is chosen to prioritize low computational latency during inference, ensuring that the model order selection remains efficient for real-time \gls{ISAC} applications. Moreover, \glspl{FNN} have been shown to be suitable for simple classification tasks \cite{doha2025}, which aligns with the objective of the proposed \gls{DL}-based approach.

\subsection{Channel Estimation Performance at Communication Receiver}
In the considered scenario, different performance metrics are evaluated to measure channel estimation performance and are listed as follows.

\subsubsection{NMSE vs Pilot SNR}  
The accuracy of channel estimation algorithms is evaluated using the normalized mean squared error (NMSE), computed as  
\begin{equation} \label{NSME}  
\text{NMSE}=\mathbb{E}\Bigg[\frac{\|\textbf{H}_{\text{DD}}-\hat{\textbf{H}}_{\text{DD}}\|^2_F}{\|\textbf{H}_{\text{DD}}\|^2_F}\Bigg].  
\end{equation}  

\begin{figure}[t]
    \centering
    \resizebox{0.99\columnwidth}{!}{
        \definecolor{mycolor1}{rgb}{0.64706,0.16471,0.16471}%
\definecolor{mycolor2}{rgb}{0.18039,0.54510,0.34118}%
\definecolor{mycolor3}{rgb}{0.49412,0.18431,0.55686}%

\begin{tikzpicture}

\begin{axis}[%
width=4.5in,
height=2.8518in,
at={(0.758in,0.529in)},
scale only axis,
xmin=0,
xmax=20,
xlabel style={font=\color{white!15!black}},
xlabel={\Large $\mathrm{SNR}_p \ \mathrm{[dB]}$},
ymin=-30,
ymax=-5,
ylabel style={font=\color{white!15!black}},
ylabel={\Large $\mathrm{NMSE} \ \mathrm{[dB]}$},
tick label style={font=\Large},
axis background/.style={fill=white},
axis x line=box, 
axis y line=box, 
xmajorgrids,
ymajorgrids,
legend style={legend cell align=left, align=left, draw=white!15!black, font=\large}
]

\addplot [color=blue, line width=2.0pt, mark=square, mark options={solid, blue, mark size=4pt}]
  table[row sep=crcr]{%
0	-8.83319907851236\\
2.5	-9.97654147276043\\
5	-11.0734373447253\\
7.5	-12.0655842202641\\
10	-12.9078344506959\\
12.5	-13.6076546395525\\
15	-14.2095951824058\\
17.5	-14.6058191936536\\
20	-14.9289941368642\\
};
\addlegendentry{Threshold Method \cite{Raviteja2019,Viterbo2022}}

\addplot [color=mycolor1, line width=2.0pt, mark=o, mark options={solid, mycolor1, mark size=4pt}]
  table[row sep=crcr]{%
0	-18.3239163943767\\
2.5	-20.8209956689816\\
5	-21.7315674723044\\
7.5	-22.4583682084198\\
10	-23.4491711782963\\
12.5	-23.4170099600026\\
15	-23.971147744487\\
17.5	-24.2551904958662\\
20	-24.3637720614075\\
};
\addlegendentry{Correlation-based, $L_h=1$}

\addplot [color=mycolor2, line width=2.6pt, mark=triangle, mark options={solid, mycolor2, mark size=4.5pt}]
  table[row sep=crcr]{%
0	-18.8937482283913\\
2.5	-21.7432658563594\\
5	-23.5577869608887\\
7.5	-24.5631634635578\\
10	-25.4758253150716\\
12.5	-26.1404758709332\\
15	-26.1665026665121\\
17.5	-26.5864201826328\\
20	-26.7270010623929\\
};
\addlegendentry{Correlation-based, $L_h=2$}

\addplot [color=mycolor3, line width=2.0pt, mark=x, mark options={solid, mycolor3, mark size=4.5pt}]
  table[row sep=crcr]{%
0	-18.9409598777184\\
2.5	-21.7573109071877\\
5	-23.5317044843692\\
7.5	-24.5548546222375\\
10	-25.5201729252883\\
12.5	-26.1209637125317\\
15	-26.2313876689201\\
17.5	-26.5987866120649\\
20	-26.7602391260757\\
};
\addlegendentry{Correlation-based, $L_h=3$}

\end{axis}
\end{tikzpicture}}
    \caption{Comparison of the threshold method \cite{Raviteja2019,Viterbo2022} and the proposed correlation-based method in Algorithm \ref{alg:CM} with prior knowledge of $P$. The normalized MSE vs pilot SNR is shown for different values of $L_h$.}
    \label{fig:NMSEvsPSNR_TMvsCM}
\end{figure}

\begin{figure}
\centering
    \resizebox{0.99\columnwidth}{!}{
        \definecolor{mycolor1}{rgb}{0.18039,0.54510,0.34118}%
\definecolor{mycolor2}{rgb}{0.92941,0.69412,0.12549}%

\begin{tikzpicture}

\begin{axis}[
width=4.5in,
height=2.8518in,
at={(0.758in,0.529in)},
scale only axis,
xmin=0,
 xmax=20,
 xlabel style={font=\color{white!15!black}},
 xlabel={\Large $\mathrm{SNR}_p \ \mathrm{[dB]}$},
 tick label style={font=\Large},
 ymin=-32,
  ymax=0,
  ylabel style={font=\color{white!15!black}},
   ylabel={\Large $\mathrm{NMSE} \ \mathrm{[dB]}$},
 axis background/.style={fill=white},
 axis x line=box,
 axis y line=box,
  xmajorgrids,
 ymajorgrids,
 legend style={at={(0.99, 0.78)},anchor=east,legend cell align=left,draw=white!15!black,font=\large}
]

\addplot [color=blue, line width=2.0pt, mark=square, mark options={solid, blue,mark size=4pt}]
  table[row sep=crcr]{%
0	-8.83319907851236\\
2.5	-9.97654147276043\\
5	-11.0734373447253\\
7.5	-12.0655842202641\\
10	-12.9078344506959\\
12.5	-13.6076546395525\\
15	-14.2095951824058\\
17.5	-14.6058191936536\\
20	-14.9289941368642\\
};
\addlegendentry{Threshold Method \cite{Raviteja2019,Viterbo2022}}

\addplot [color=cyan, line width=2.0pt, mark=diamond, mark options={solid, cyan,mark size=4pt}]
  table[row sep=crcr]{%
0	-8.07344\\
2.5	-8.78727\\
5	-10.9484\\
7.5	-13.4727\\
10	-16.2731\\
12.5	-21.4525\\
15	-23.9144\\
17.5	-27.8476\\
20	-31.372\\
};
\addlegendentry{PIPIC \cite{marchese2024}, SC}

\addplot [color=violet, line width=2.0pt, mark=star, mark options={solid, violet,mark size=4pt}]
  table[row sep=crcr]{%
0	-4.49\\
2.5	-7.1\\
5	-10.014\\
7.5	-12.87\\
10	-15.59\\
12.5	-17.52\\
15	-19.3\\
17.5	-21.32\\
20	-23.369\\
};
\addlegendentry{TSE \cite{Khan2021,Khan2023}, SC}

\addplot [color=mycolor1, line width=2.0pt, mark=triangle, mark options={solid, mycolor1,mark size=4pt}]
  table[row sep=crcr]{%
0	-18.8937482283913\\
2.5	-21.7432658563594\\
5	-23.5577869608887\\
7.5	-24.5631634635578\\
10	-25.4758253150716\\
12.5	-26.1404758709332\\
15	-26.1665026665121\\
17.5	-26.5864201826328\\
20	-26.7270010623929\\
};
\addlegendentry{Correlation-based, $L_h=2$}

\addplot [color=mycolor2, line width=2.0pt, mark=o, mark options={solid, mycolor2,mark size=4pt}]
  table[row sep=crcr]{%
0	-15.0619939306773\\
2.5	-18.04568205641\\
5	-21.1919705950657\\
7.5	-21.817695901594\\
10	-23.308093158337\\
12.5	-23.3248379625417\\
15	-24.610874860266\\
17.5	-24.178591279016\\
20	-24.9084359604984\\
};
\addlegendentry{Correlation-based, $L_h=2$, FNN}

\addplot [color=red, line width=2.0pt, mark=x, mark options={solid, red,mark size=4pt}]
  table[row sep=crcr]{%
0	-4.28700525547063\\
2.5	-7.19845410001938\\
5	-10.1726087081143\\
7.5	-12.6026486102399\\
10	-16.8740325669041\\
12.5	-20.9638622386165\\
15	-23.452695695076\\
17.5	-26.9363007267447\\
20	-28.1197633134946\\
};
\addlegendentry{Correlation-based, $L_h=2$, SC}

\end{axis}
\end{tikzpicture}%
}
    \caption{Comparison of the proposed \gls{FNN}-based estimation of $P$ against the conventional \gls{SC}-based approach. The normalized MSE vs pilot SNR is shown for the optimum number of hierarchical levels $L_h=2$.} 
    \label{fig:NMSEvsPSNR_FNNvsSC} 
\end{figure}

Fig.~\ref{fig:NMSEvsPSNR_TMvsCM} illustrates the \gls{NMSE} as a function of pilot \gls{SNR} for the proposed method in Algorithm \ref{alg:CM}, considering different numbers of hierarchical levels ($L_h=1,2,3$) with prior knowledge of $P$. As a baseline, the conventional threshold method \cite{Raviteja2019,Viterbo2022} is also reported. As expected, the performance of the baseline method is limited by \gls{IPI}, achieving a minimum \gls{NMSE} of $-15$ dB at $\text{SNR}_p=20$ dB. Conversely, the proposed correlation-based method attains a lower \gls{NMSE}. Specifically, when $L_h=1$, the proposed method provides a performance gain of nearly $10$ dB over the entire $\text{SNR}_p$ range compared to the baseline method. Increasing the number of hierarchical levels to $L_h=2$ yields an additional gain of approximately $2$ dB at sufficiently high $\text{SNR}_p$. However, using $L_h=3$ does not result in further performance improvements beyond those obtained with $L_h=2$, making $L_h=2$ the optimal choice. Instead, selecting $L_h=3$ increases computational complexity without enhancing estimation accuracy.  

Fig.~\ref{fig:NMSEvsPSNR_FNNvsSC} shows the \gls{NMSE} as a function of pilot \gls{SNR} when estimating $P$. The proposed \gls{FNN}-based approach in \eqref{eq:Pestimation} is compared with the \gls{SC}-based approach (specifically, the \gls{TSE} method proposed in \cite{Khan2021,Khan2023} performing disjoint \gls{DD} estimation and the \gls{PIPIC} proposed in \cite{marchese2024} that performs joint \gls{DD} estimation). Additionally, the performance of the proposed approach with \gls{SC}-based estimation of $P$ is reported. The threshold method and the proposed correlation-based method with prior knowledge of $P$ are included as upper and lower benchmarks for comparison.  
As expected, due to the generalization capabilities of the neural network, the proposed \gls{FNN}-based approach significantly outperforms the \gls{SC}-based approach at low $\text{SNR}_p$, providing more stable estimation accuracy across different $\text{SNR}_p$ values and exhibiting a performance loss of only $2$-$3$ dB compared to the ideal case where $P$ is known. Furthermore, for $\text{SNR}_p$ in the range $[0,6]$ dB, the \gls{SC}-based approach underperforms the threshold method. This is because, at very low $\text{SNR}_p$, the \gls{SC}-based approach fails to detect lower-amplitude paths that are masked by the \gls{SC}, allowing the simpler threshold method to achieve a lower estimation error. On the other hand, at high $\text{SNR}_p$, \gls{SC}-based approaches achieve better performance at the cost of overestimating $P$. Additionally, in mid-range $\text{SNR}_p$ conditions, the performance of PIPIC is nearly identical to that of the proposed approach with \gls{SC}-based estimation of $P$. Thus, in the proposed approach, disjoint estimation reduces complexity without sacrificing performance. On the other hand, performance of the \gls{TSE}, which also performs disjoint \gls{DD} estimation, is lower than that of the \gls{PIPIC} and the proposed correlation method. Specifically, at high \gls{SNR} the performance gain of the proposed correlation based method against the \gls{TSE} is about $4$-$5$ dB of \gls{NMSE}. This performance gain arises because the disjoint \gls{DD} estimation in the \gls{TSE} algorithm relies on an approximate \gls{ML} formulation that is valid only for large \gls{OTFS} frames \cite{Khan2021,Khan2023}. In contrast, our proposed method leverages an exact algebraic separability that remains robust even in scenarios with significant \gls{IPI}.

\subsubsection{NMSE vs OTFS Frame Size}  
The estimation accuracy varies as the number of delay-Doppler bins changes. Given that a low value of $N$ is used in the considered scenario, it is particularly relevant to evaluate estimation accuracy as a function of the number of time slots $N$.  

Fig.~\ref{fig:NMSEvsN} presents the NMSE as a function of the number of Doppler bins for both the threshold method \cite{Raviteja2019,Viterbo2022} and the proposed correlation-based method. Simulations are conducted with a fixed $M=64$ and at $\text{SNR}_p=15$ dB. It can be observed that, since the threshold method is primarily limited by IPI, increasing $N$ from $8$ to $64$ provides a maximum performance gain of $3$ dB. Conversely, the proposed correlation-based method accounts for IPI, resulting in a performance gain of $6$ dB when increasing $N$. Furthermore, the proposed correlation-based method achieves an improvement of nearly $15$ dB compared to the baseline method when $M=N=64$.  

\begin{figure}
\centering
    \resizebox{0.99\columnwidth}{!}{
%
%
\definecolor{mycolor1}{rgb}{0.18039,0.54510,0.34118}%
\begin{tikzpicture}

\begin{axis}[%
width=4.521in,
height=2in,
at={(0.758in,0.528in)},
scale only axis,
xmin=8,
xmax=64,
xlabel style={font=\color{white!15!black}},
xlabel={\Large $N$},
ymin=-35,
ymax=-10,
ylabel style={font=\color{white!15!black}},
ylabel={\Large $\mathrm{NMSE} \ \mathrm{[dB]}$},
tick label style={font=\Large},
axis background/.style={fill=white},
axis x line=box,
axis y line=box,
xmajorgrids,
ymajorgrids,
legend style={at={(0.99,0.37)}, anchor=south east,legend cell align=left, align=left, draw=white!15!black, font=\Large}
]
\addplot [color=blue, line width=2.0pt, mark=square, mark options={solid, blue, mark size=4pt}]
  table[row sep=crcr]{%
8	-13.3874668245709\\
16	-14.2241960884025\\
32	-15.3648424763842\\
64	-16.0926643231426\\
};
\addlegendentry{Threshold Method \cite{Raviteja2019,Viterbo2022}}

\addplot [color=mycolor1, line width=2.0pt, mark=triangle, mark options={solid, mycolor1,mark size=4pt}]
  table[row sep=crcr]{%
8	-24.0167644164868\\
16	-25.7739544410057\\
32	-30.0311590278676\\
64	-30.947330488538\\
};
\addlegendentry{Correlation-based, $L_h=2$}

\end{axis}
\end{tikzpicture}%
}
    \caption{Channel estimation accuracy measured through the NMSE is shown as a function of the number of Doppler bins. Both the baseline threshold method and the proposed correlation-based method with $L_h=2$ are shown.} 
    \label{fig:NMSEvsN} 
\end{figure}

\subsubsection{Model Order Selection}  
As the proposed approach uses the \gls{FNN} for model order selection, a comparison of the average number of detected paths for both the baseline \gls{SC}-based approach and the proposed \gls{FNN}-based approach is illustrated in Fig.~\ref{fig:Pavg}. Specifically, the figure shows the average number of detected paths as a function of $\text{SNR}_p$. As expected, the neural network is able to accurately estimate the correct number of paths on average. In contrast, the \gls{SC}-based approach significantly underestimates $P$ at low $\text{SNR}_p$. Therefore, the proposed \gls{FNN}-based approach effectively improves estimation performance in low $\text{SNR}_p$ conditions.  

\begin{figure}[t]
\centering
    \resizebox{0.99\columnwidth}{!}{
%
%
\definecolor{mycolor1}{rgb}{0.92941,0.69412,0.12549}%
\definecolor{mycolor2}{rgb}{1.00000,1.00000,0.00000}%
\begin{tikzpicture}

\begin{axis}[%
width=4.521in,
height=2.356in,
at={(0.758in,0.481in)},
scale only axis,
xmin=0,
xmax=20,
xlabel style={font=\color{white!15!black}},
xlabel={\Large $\mathrm{SNR}_p \ \mathrm{[dB]}$},
ymin=0,
ymax=5,
ylabel style={font=\color{white!15!black}},
ylabel={\Large Average model order selection},
tick label style={font=\Large},
axis background/.style={fill=white},
axis x line=box,
axis y line=box,
xmajorgrids,
ymajorgrids,
legend pos=south east,
legend style={legend cell align=left, align=left, draw=white!15!black,  font=\Large}
]

\addplot [color=mycolor1, line width=1.5pt, mark=o, mark options={solid, mycolor1, mark size=4pt}]
  table[row sep=crcr]{%
0	3.55\\
2.5	3.8\\
5	3.8\\
7.5	3.83\\
10	3.88\\
12.5	3.9\\
15	3.86\\
17.5	3.88\\
20	3.84\\
};
\addlegendentry{FNN}

\addplot[area legend, draw=black, fill=mycolor2, fill opacity=0.3]
table[row sep=crcr] {%
x	y\\
0	4.43048425610713\\
2.5	4.66456621925376\\
5	4.48164981086073\\
7.5	4.54145036099619\\
10	4.53566682842659\\
12.5	4.51134063641915\\
15	4.4959594676113\\
17.5	4.45348835113618\\
20	4.47117655088243\\
20	3.20882344911757\\
17.5	3.30651164886382\\
15	3.2240405323887\\
12.5	3.28865936358085\\
10	3.2243331715734\\
7.5	3.11854963900381\\
5	3.11835018913927\\
2.5	2.93543378074624\\
0	2.66951574389287\\
}--cycle;
\addlegendentry{$\pm\sigma_P$, FNN}

\addplot [color=red, line width=1.5pt, mark=x, mark options={solid, red,mark size=4pt}]
  table[row sep=crcr]{%
0	1.2\\
2.5	1.76\\
5	2.32\\
7.5	2.77\\
10	3.14\\
12.5	3.42\\
15	3.63\\
17.5	3.86\\
20	4\\
};
\addlegendentry{SC}

\addplot[area legend, draw=black, fill=red, fill opacity=0.3]
table[row sep=crcr] {%
x	y\\
0	1.60201512610368\\
2.5	2.38150030270518\\
5	3.02896139713405\\
7.5	3.50656313546447\\
10	3.85095326530971\\
12.5	4.05849570685853\\
15	4.15522241560941\\
17.5	4.20873508801978\\
20	4.37605071654518\\
20	3.62394928345482\\
17.5	3.51126491198022\\
15	3.10477758439059\\
12.5	2.78150429314147\\
10	2.42904673469029\\
7.5	2.03343686453553\\
5	1.61103860286595\\
2.5	1.13849969729482\\
0	0.797984873896315\\
}--cycle;
\addlegendentry{$\pm\sigma_P$, SC}

\addplot [color=black, dashed, line width=1.5pt]
  table[row sep=crcr]{%
0	4\\
2.5	4\\
5	4\\
7.5	4\\
10	4\\
12.5	4\\
15	4\\
17.5	4\\
20	4\\
};
\addlegendentry{True $P$}

\end{axis}
\end{tikzpicture}%
}
    \caption{Model order selection comparison between the conventional SC-based approach against the proposed FNN-based approach. The true value of $P$ is also shown for the comparison.}
    \label{fig:Pavg} 
\end{figure}

\begin{figure}[t]
\centering
    \resizebox{0.99\columnwidth}{!}{
%
%
\definecolor{mycolor1}{rgb}{0.92941,0.69412,0.12549}%
\definecolor{mycolor2}{rgb}{1.00000,1.00000,0.00000}%
\begin{tikzpicture}

\begin{axis}[%
width=4.521in,
height=2.356in,
at={(1.212in,0.624in)},
scale only axis,
xmin=0,
xmax=4,
xlabel style={font=\color{white!15!black}},
xlabel={\Large Normalized delay distance, $\Delta l$},
ymin=1.8,
ymax=3.3,
ylabel style={font=\color{white!15!black}},
ylabel={\Large Average model order selection},
tick label style={font=\Large},
axis background/.style={fill=white},
axis x line=box,
axis y line=box,
xmajorgrids,
ymajorgrids,
legend style={
    at={(0.97, 0.2)},
    anchor=south east,
    legend cell align=left, 
    align=left, 
    draw=white!15!black, 
    font=\large
}
]
\addplot [color=mycolor1, line width=2.0pt, mark=o, mark options={solid, mycolor1, mark size=4pt}]
  table[row sep=crcr]{%
4	3.106\\
3.5	3.091\\
3	3.133\\
2.5	2.884\\
2	2.861\\
1.5	2.787\\
1	2.252\\
0.5	2.044\\
0	2.067\\
};
\addlegendentry{FNN, SNR$=5$ dB}

\addplot [dashed, color=mycolor1, line width=2.0pt, mark=o, mark options={solid, mycolor1, mark size=4pt}]
  table[row sep=crcr]{%
4	3.098\\
3.5	3.1\\
3	3.105\\
2.5	2.931\\
2	2.915\\
1.5	2.815\\
1	2.372\\
0.5	2.042\\
0	2.116\\
};
\addlegendentry{FNN, SNR$=18$ dB}

\addplot [color=red, line width=2.0pt, mark=x, mark options={solid, red,mark size=4pt}]
  table[row sep=crcr]{%
4	1.99666666666667\\
3.5	2.02333333333333\\
3	1.99\\
2.5	2.05666666666667\\
2	2.03666666666667\\
1.5	2.01333333333333\\
1	2.00333333333333\\
0.5	2\\
0	1.93333333333333\\
};
\addlegendentry{SC, SNR$=5$ dB}

\addplot [dashed, color=red, line width=2.0pt, mark=x, mark options={solid, red,mark size=4pt}]
  table[row sep=crcr]{%
4	2.92666666666667\\
3.5	2.92333333333333\\
3	2.96666666666667\\
2.5	2.94333333333333\\
2	3.01666666666667\\
1.5	2.99333333333333\\
1	3.22666666666667\\
0.5	3.15666666666667\\
0	3.01666666666667\\
};
\addlegendentry{SC, SNR$=18$ dB}

\addplot [color=black, dashed, line width=2pt]
  table[row sep=crcr]{%
4	3\\
3.5	3\\
3	3\\
2.5	3\\
2	3\\
1.5	3\\
1	3\\
0.5	3\\
0	3\\
};
\addlegendentry{True $P$}

\end{axis}

\end{tikzpicture}%
}
    \caption{Average number of detected paths versus delay distance for the SC-based approach and the proposed FNN-based method at different SNR levels ($\text{SNR}_p = 5$ dB and $\text{SNR}_p = 18$ dB). The dashed line represents the true value of $P$.}
    \label{fig:ResFNN} 
\end{figure}

\subsubsection{FNN Resolution Analysis}
In this section, the resolution capabilities of the proposed \gls{FNN}-based model order selection are investigated. Fig. \ref{fig:ResFNN} illustrates the average number of detected paths as a function of the delay distance between paths for both the \gls{SC}-based approach and the proposed \gls{FNN} architecture. The scenario includes $P=3$ paths, where the first path delay is kept fixed ($l_1=0$), while the second and third paths are positioned to test the resolution limits of the estimator. Specifically, the second path delay is set to $l_2 = \tau_2 / \Delta\tau$ (with $\tau_2 = 2.4$ $\mu$s), and the third path is placed at $l_3 = l_2 + \Delta l$, where $\Delta l$ is a normalized delay distance. The \gls{PDP} is defined as $[0, -1, -5]$ dB, resulting in a $4$ dB dynamic range between the third and second paths. 
As observed in the simulation results, the \gls{SC}-based approach suffers from significant path underestimation at low \gls{SNR} ($\text{SNR}_p = 5$ dB), regardless of the delay distance, as the weaker path is often masked by noise and \gls{IPI}. Conversely, at high \gls{SNR} ($\text{SNR}_p = 18$ dB), the \gls{SC}-based method tends to overestimate the number of paths when the delay distance is small due to the \gls{IPI} that creates residual peaks in the \gls{DD} domain.
In contrast, the \gls{FNN}-based approach demonstrates superior robustness. Even at low \gls{SNR}, the neural network accurately identifies the correct number of paths as the delay distance increases beyond a certain threshold. The generalization capabilities of the \gls{FNN} allow it to effectively distinguish between closely spaced paths and noise-induced artifacts, providing a more stable model order selection under varying \gls{SNR} conditions. These results further highlight that the proposed \gls{DL}-aided architecture enhances path detection performance in challenging high-\gls{IPI} and low-\gls{SNR} scenarios.

\begin{figure}[t]
    \centering
    \resizebox{0.99\columnwidth}{!}{\definecolor{mycolor1}{rgb}{0.18039,0.54510,0.34118}%
\definecolor{mycolor2}{rgb}{0.92941,0.69412,0.12549}%
\definecolor{mycolor3}{rgb}{0.8500, 0.3250, 0.0980}

\begin{tikzpicture}

\begin{axis}[
width=4.5in,
height=2.3518in,
at={(0.758in,0.529in)},
scale only axis,
xmin=0,
 xmax=20,
 xlabel style={font=\color{white!15!black}},
 xlabel={\Large $\mathrm{SNR}_p \ \mathrm{[dB]}$},
 tick label style={font=\Large},
 ymin=-25,
  ymax=0,
  ylabel style={font=\color{white!15!black}},
   ylabel={\Large $\mathrm{NMSE} \ \mathrm{[dB]}$},
 axis background/.style={fill=white},
 axis x line=box,
 axis y line=box,
  xmajorgrids,
 ymajorgrids,
 legend style={at={(0.98, 0.8)},anchor=east,legend cell align=left,draw=white!15!black,font=\large}
]

\addplot [color=mycolor2, line width=2.0pt, mark=o, mark options={solid, mycolor2,mark size=4pt}]
  table[row sep=crcr]{%
0	-11.5441337909316\\
2.5	-13.1848771653476\\
5	-12.6296608977686\\
7.5	-13.6104856322695\\
10	-13.065137652314\\
12.5	-13.6784746639539\\
15	-13.1235647476469\\
17.5	-12.8487043272666\\
20	-13.2185182489574\\
};
\addlegendentry{FNN}

\addplot [color=mycolor3, line width=2.0pt, mark=diamond, mark options={solid, mycolor3,mark size=4pt}]
  table[row sep=crcr]{%
0	-12.7241741783028\\
2.5	-13.4310400143756\\
5	-14.5106044404655\\
7.5	-15.5936604485546\\
10	-17.331161135889\\
12.5	-17.6798359472943\\
15	-18.4627768447422\\
17.5	-17.6005101249929\\
20	-17.9004055023424\\
};
\addlegendentry{$\text{FNN}_{\text{2-8}}$}

\addplot [color=red, line width=2.0pt, mark=x, mark options={solid, red,mark size=4pt}]
  table[row sep=crcr]{%
0	-3.52142140153284\\
2.5	-6.5318881285174\\
5	-8.80126736990922\\
7.5	-12.0837929725836\\
10	-15.2157217612897\\
12.5	-17.9715228617951\\
15	-21.2633600751546\\
17.5	-22.8843907094859\\
20	-23.7802944208829\\
};
\addlegendentry{SC}

\end{axis}
\end{tikzpicture}%
}
    \caption{Comparison of the proposed FNN-based model order selection scheme against the SC-based approach. The NMSE performance is shown against pilot SNR using the proposed correlation-based channel estimation algorithm with $L_h=2$. Finally, FNN denotes the DL scheme, trained as discussed in Section \ref{sec:FNNtraining}; while $\text{FNN}_{\text{2-8}}$ denotes the same FNN trained with a dataset including channels with $P\in[2,8]$.}
    \label{fig:NMSEvsPSNR_6paths}
\end{figure}

\begin{figure}[t]
\centering
    \resizebox{0.99\columnwidth}{!}{
%
%
\definecolor{mycolor1}{rgb}{0.92941,0.69412,0.12549}%
\definecolor{mycolor2}{rgb}{1.00000,1.00000,0.00000}%
\definecolor{mycolor3}{rgb}{0.8500, 0.3250, 0.0980}%

\begin{tikzpicture}

\begin{axis}[%
width=4.521in,
height=2.356in,
at={(0.758in,0.481in)},
scale only axis,
xmin=0,
xmax=20,
xlabel style={font=\color{white!15!black}},
xlabel={\Large $\mathrm{SNR}_p \ \mathrm{[dB]}$},
ymin=1,
ymax=7,
ylabel style={font=\color{white!15!black}},
ylabel={\Large Average model order selection},
tick label style={font=\Large},
axis background/.style={fill=white},
axis x line=box,
axis y line=box,
xmajorgrids,
ymajorgrids,
legend pos=south east,
legend style={legend cell align=left, align=left, draw=white!15!black,  font=\large}
]

\addplot [color=mycolor1, line width=1.5pt, mark=o, mark options={solid, mycolor1, mark size=4pt}]
  table[row sep=crcr]{%
0	4.65\\
2.5	4.9\\
5	4.8\\
7.5	4.9\\
10	4.85\\
12.5	4.9\\
15	4.85\\
17.5	4.75\\
20	4.85\\
};
\addlegendentry{FNN}

\addplot[area legend, draw=black, fill=mycolor2, fill opacity=0.3]
table[row sep=crcr] {%
x	y\\
0	5.2371429486124\\
2.5	5.20779350562555\\
5	5.3231483637806\\
7.5	5.20779350562555\\
10	5.33936048492959\\
12.5	5.20779350562555\\
15	5.21634754853252\\
17.5	5.19426165831932\\
20	5.21634754853252\\
20	4.48365245146748\\
17.5	4.30573834168068\\
15	4.48365245146748\\
12.5	4.59220649437445\\
10	4.36063951507041\\
7.5	4.59220649437445\\
5	4.2768516362194\\
2.5	4.59220649437445\\
0	4.0628570513876\\
}--cycle;
\addlegendentry{$\pm\sigma_P$, FNN}

\addplot [color=mycolor3, line width=1.5pt, mark=diamond, mark options={solid, mycolor3, mark size=4pt}]
  table[row sep=crcr]{%
0	5.34\\
2.5	5.56\\
5	5.74\\
7.5	5.8\\
10	6.04\\
12.5	5.98\\
15	6.1\\
17.5	6.04\\
20	5.92\\
};
\addlegendentry{$\text{FNN}_{\text{2-8}}$}

\addplot[area legend, draw=black, fill=mycolor3, fill opacity=0.15]
table[row sep=crcr] {%
x	y\\
0	6.84658418894111\\
2.5	7.12700676580595\\
5	7.24929095394939\\
7.5	7.21421356237309\\
10	7.28474405204526\\
12.5	7.11371342165866\\
15	7.24731274315779\\
17.5	7.2173370158878\\
20	7.05999641961486\\
20	4.78000358038514\\
17.5	4.8626629841122\\
15	4.95268725684221\\
12.5	4.84628657834134\\
10	4.79525594795474\\
7.5	4.3857864376269\\
5	4.23070904605061\\
2.5	3.99299323419405\\
0	3.83341581105889\\
}--cycle;
\addlegendentry{$\pm\sigma_P$, $\text{FNN}_{\text{2-8}}$}

\addplot [color=red, line width=1.5pt, mark=x, mark options={solid, red,mark size=4pt}]
  table[row sep=crcr]{%
0	1.7\\
2.5	2.9\\
5	3.6\\
7.5	4.45\\
10	5.15\\
12.5	5.45\\
15	5.85\\
17.5	5.95\\
20	6.2\\
};
\addlegendentry{SC}

\addplot[area legend, draw=black, fill=red, fill opacity=0.3]
table[row sep=crcr] {%
x	y\\
0	2.27124057057748\\
2.5	3.54072327551719\\
5	4.28055704737872\\
7.5	4.96041778553404\\
10	5.7371429486124\\
12.5	6.0548053188293\\
15	6.21634754853252\\
17.5	6.17360679774998\\
20	6.7231483637806\\
20	5.6768516362194\\
17.5	5.72639320225002\\
15	5.48365245146748\\
12.5	4.8451946811707\\
10	4.5628570513876\\
7.5	3.93958221446596\\
5	2.91944295262128\\
2.5	2.25927672448281\\
0	1.12875942942252\\
}--cycle;
\addlegendentry{$\pm\sigma_P$, SC}

\addplot [color=black, dashed, line width=1.5pt]
  table[row sep=crcr]{%
0	6\\
2.5	6\\
5	6\\
7.5	6\\
10	6\\
12.5	6\\
15	6\\
17.5	6\\
20	6\\
};
\addlegendentry{True $P$}

\end{axis}
\end{tikzpicture}%
}
    \caption{Model order selection comparison between the conventional SC-based approach against the proposed FNN-based approach in a channel with $P=6$ paths. Here, $\text{FNN}_{\text{2-8}}$ denotes the same FNN trained with a dataset including channels with $P\in[2,8]$.}
    \label{fig:Pavg_6paths} 
\end{figure}

\subsubsection{FNN Performance Under Increased Model Order}

Since, as described in Section \ref{sec:FNNtraining}, the \gls{FNN} scheme has been trained on a limited dataset, in Fig. \ref{fig:NMSEvsPSNR_6paths} and \ref{fig:Pavg_6paths} channel estimation and model order selection performance is investigated in a channel with $P=6$ paths to analyze the impact of \gls{FNN} behavior when faced with an unseen class. Since the model is constrained by its final layer to a fixed set of outcomes, the estimated number of paths is expected to saturate at the maximum trained cardinality (i.e., $\hat{P}=5$), as the network cannot infer a new output class beyond its predefined architectural limits.
This behavior is highlighted in Fig. \ref{fig:Pavg_6paths}, where it can be noted that the average number of paths detected by the \gls{FNN} is around $5$, regardless of the \gls{SNR}. On the other hand, the \gls{SC}-based approach exhibits \gls{SNR}-dependent model order selection capabilities, underestimating the number of paths at low \gls{SNR}. Fig. \ref{fig:Pavg_6paths} shows also the average number of paths detected by the same \gls{FNN} described in Section \ref{sec:FNNtraining} and trained with a dataset with the same size (18000 samples) but with a larger number of classes, specifically $P\sim\mathcal{U}[2,8]$. It can be observed that the model order selection capabilities are increased as the \gls{FNN} is actually capable of detecting the correct number of paths on average. On the other hand, it can be noted that the variance of the estimate is increased since the same dataset size has been used for a larger number of classes.
These model order selection capabilities affect the channel estimation error, as depicted in Fig. \ref{fig:NMSEvsPSNR_6paths}, where it can be noted that at low \gls{SNR} the \gls{FNN}-based approach allows achieving lower \gls{NMSE}. Moreover, the use of the same \gls{FNN}, trained with a large number of classes leads to a \gls{NMSE} gain of almost $5$ dB at high \gls{SNR}.

\begin{figure}[t]
    \centering
    \resizebox{0.99\columnwidth}{!}{
%
%
\definecolor{mycolor1}{rgb}{0.92941,0.69412,0.12549}%
\definecolor{mycolor2}{rgb}{1.00000,1.00000,0.00000}%
\definecolor{mycolor3}{rgb}{0.8500, 0.3250, 0.0980}%

\begin{tikzpicture}

\begin{axis}[%
width=4.221in,
height=2.356in,
scale only axis,
xmin=-10,
xmax=30,
xlabel={\Large Rician K-factor [dB]},
ymin=-26,
ymax=-8,
tick label style={font=\Large},
ylabel={\Large NMSE [dB]},
axis background/.style={fill=white},
xmajorgrids,
ymajorgrids,
legend style={legend cell align=left, align=left}
]
\addplot [color=mycolor1, line width=2.0pt, mark=o, mark options={solid, mycolor1,mark size=4pt}]
  table[row sep=crcr]{%
-10	-21.4973071587315\\
-5	-20.598098719759\\
0	-20.8277020639438\\
5	-20.3094267885552\\
10	-19.3908740826289\\
15	-18.5920080316209\\
20	-19.8469530592939\\
25	-22.2058582496854\\
30	-24.1780016686898\\
};
\addlegendentry{\Large FNN}

\addplot [color=red, line width=2.0pt, mark=x, mark options={solid, red,mark size=4pt}]
  table[row sep=crcr]{%
-10	-10.0100396926105\\
-5	-10.4164650119923\\
0	-10.7556099987134\\
5	-10.0712174911492\\
10	-9.61760976442071\\
15	-11.0272303101436\\
20	-15.9847071899303\\
25	-19.915584019149\\
30	-24.6485596742972\\
};
\addlegendentry{\Large SC}

\end{axis}

\end{tikzpicture}%
}
    \caption{Comparison of the proposed FNN-based model order selection scheme against the SC-based approach under Rician fading. The NMSE performance is shown against Rician $K$ factor using the proposed correlation-based channel estimation algorithm with $L_h=2$.}
    \label{fig:NMSEvsRicianK}
\end{figure}

\subsubsection{NMSE vs Rician $K$-factor}
The robustness of the proposed \gls{FNN}-based channel estimation method against varying propagation conditions is investigated by considering a Rician fading channel. Specifically, the impact of a predominant \gls{LoS} component on the \gls{NMSE} is evaluated. 
For this analysis, the \gls{SNR} is fixed at $\mathrm{SNR}_p = 5$ dB to consider a low-\gls{SNR} regime, while the Rician $K$-factor is varied from $-10$ dB to $30$ dB. The channel model consists of $P=4$ propagation paths, where the delay and Doppler shifts are generated according to the methodology described in Section \ref{sec:SimScenario}. The channel coefficients are modeled such that the first path ($i=1$) contains the \gls{LoS} component. Specifically, the $i$-th coefficient is expressed as
\begin{equation}
g_i = 
\begin{cases} 
\sqrt{\frac{K}{K+1}} e^{j\theta} +  w_i, & i=1 \\
w_i, & i > 1
\end{cases},
\end{equation}
where $w_i \sim \mathcal{CN}(0, \frac{1}{K+1})$ represents the diffuse scattering component and $\theta\sim\mathcal{U}[0,2\pi]$ is the \gls{LoS} phase. Finally, channel gains are normalized according to Section \ref{sec:PilotModel}.
Fig. \ref{fig:NMSEvsRicianK} shows the \gls{NMSE} against the Rician $K$ factor for the proposed correlation based channel estimation algorithm. It can be noted that, the \gls{SC}-based scheme achieves low \gls{NMSE} when the \gls{LoS} component is dominant, as the \gls{LoS} is correctly identified and the diffuse scattering component are typically missed, since they fall below the noise level. On the other hand, when the channel starts becoming Rayleigh faded (as $K$ decreases), performance of the \gls{SC}-based algorithm drops and the \gls{NMSE} increases to $-10$ dB. Conversely, the \gls{FNN}-based scheme achieves much lower and stable \gls{NMSE} across the entire range of $K$, with a performance gain of about $10$ dB for lower values of $K$.

\begin{figure}[t]
\centering
    \resizebox{0.99\columnwidth}{!}{
%
%
\definecolor{mycolor1}{rgb}{0.00000,0.44700,0.74100}%
\definecolor{mycolor2}{rgb}{0.85000,0.32500,0.09800}%
\definecolor{mycolor3}{rgb}{0.92900,0.69400,0.12500}%
\begin{tikzpicture}

\begin{axis}[%
width=4.521in,
height=2.8509in,
at={(0.758in,0.528in)},
scale only axis,
xmin=-20,
xmax=20,
xlabel style={font=\color{white!15!black}},
xlabel={\Large $\mathrm{SNR}_{\text{rad}} \ \mathrm{[dB]}$},
ymode=log,
ymin=1e-1,
ymax=100,
yminorticks=true,
ylabel style={font=\color{white!15!black}},
ylabel={\Large $\mathrm{RMSE \ range} \ \mathrm{[m]}$},
tick label style={font=\Large},
axis background/.style={fill=white},
axis x line=box,
axis y line=box,
xmajorgrids,
ymajorgrids,
yminorgrids,
legend style={at={(0.03,0.03)}, anchor=south west, legend cell align=left, align=left, draw=white!15!black, font=\large}
]
\addplot [color=mycolor1, line width=2.0pt, mark=o, mark options={solid, mycolor1,mark size=4pt}]
  table[row sep=crcr]{%
-20	58.7826737799884\\
-15	15.4278635957311\\
-10	12.5056759352739\\
-5	10.8441840524122\\
0	10.7507398969975\\
5	10.6564763830825\\
10	10.2707746146239\\
15	10.0221760416667\\
20	10.0221760416667\\
};
\addlegendentry{Correlation-based, $L_h=1$}

\addplot [color=mycolor2, line width=2.0pt, mark=triangle, mark options={solid, mycolor2,mark size=4pt}]
  table[row sep=crcr]{%
-20	59.0351562334725\\
-15	12.9421634591839\\
-10	9.21599515131673\\
-5	4.57763452350289\\
0	2.61228105984929\\
5	1.75550097119582\\
10	1.06892054541749\\
15	0.669438512902593\\
20	0.619883574940669\\
};
\addlegendentry{Correlation-based, $L_h=2$}

\addplot [color=mycolor3, line width=2.0pt, mark=x, mark options={solid, mycolor3,mark size=4pt}]
  table[row sep=crcr]{%
-20	59.0043966167718\\
-15	13.0460641564896\\
-10	9.09487553826764\\
-5	4.55011516507827\\
0	2.56647637449244\\
5	1.68778642126614\\
10	0.82358048097643\\
15	0.468708181741324\\
20	0.27806557428347\\
};
\addlegendentry{Correlation-based, $L_h=3$}

\addplot [color=cyan, line width=2.0pt, mark=diamond, mark options={solid, cyan, mark size=4pt}]
  table[row sep=crcr]{%
-20	182.105\\
-15	16.4809\\
-10	8.99613\\
-5	4.21126\\
0	2.52767\\
5	1.62531\\
10	0.88884\\
15	0.495406\\
20	0.252274\\
};
\addlegendentry{PIPIC \cite{marchese2024}}

\addplot [color=black, dashed, line width=2.0pt]
  table[row sep=crcr]{%
-20	28.7820982178297\\
-15	16.1853632535731\\
-10	9.1016986207015\\
-5	5.11826126384845\\
0	2.87820982178297\\
5	1.61853632535731\\
10	0.91016986207015\\
15	0.511826126384845\\
20	0.287820982178297\\
};
\addlegendentry{$\sqrt{\text{CRLB}(d)}$}

\end{axis}
\end{tikzpicture}%
}
    \resizebox{0.99\columnwidth}{!}{
%
%
\definecolor{mycolor1}{rgb}{0.00000,0.44700,0.74100}%
\definecolor{mycolor2}{rgb}{0.85000,0.32500,0.09800}%
\definecolor{mycolor3}{rgb}{0.92900,0.69400,0.12500}%
\begin{tikzpicture}

\begin{axis}[%
width=4.521in,
height=2.8509in,
at={(0.758in,0.528in)},
scale only axis,
xmin=-20,
xmax=20,
xlabel style={font=\color{white!15!black}},
xlabel={\Large $\mathrm{SNR}_{\text{rad}} \ \mathrm{[dB]}$},
ymode=log,
ymin=5*1e-3,
ymax=3,
yminorticks=true,
ylabel style={font=\color{white!15!black}},
ylabel={\Large $\mathrm{RMSE \ velocity} \ \mathrm{[m/s]}$},
tick label style={font=\Large},
axis background/.style={fill=white},
axis x line=box,
axis y line=box,
xmajorgrids,
ymajorgrids,
yminorgrids,
legend style={at={(0.03,0.03)}, anchor=south west, legend cell align=left, align=left, draw=white!15!black, font=\large}
]
\addplot [color=mycolor1, line width=2.0pt, mark=o, mark options={solid, mycolor1,mark size=4pt}]
  table[row sep=crcr]{%
-20	1.87963189868642\\
-15	0.737783906541094\\
-10	0.506979733188298\\
-5	0.476539549150943\\
0	0.4060889953739\\
5	0.386474908277993\\
10	0.372826022569441\\
15	0.372826022569441\\
20	0.372826022569441\\
};
\addlegendentry{Correlation-based, $L_h=1$}

\addplot [color=mycolor2, line width=2.0pt, mark=triangle, mark options={solid, mycolor2,mark size=4pt}]
  table[row sep=crcr]{%
-20	1.86592390096279\\
-15	0.620606160088842\\
-10	0.356042524607951\\
-5	0.223485879426257\\
0	0.131502098972176\\
5	0.0760125695189946\\
10	0.0397905777996686\\
15	0.0310873109861971\\
20	0.0317423287885535\\
};
\addlegendentry{Correlation-based, $L_h=2$}

\addplot [color=mycolor3, line width=2.0pt, mark=x, mark options={solid, mycolor3,mark size=4pt}]
  table[row sep=crcr]{%
-20	1.8655348687244\\
-15	0.623986175570528\\
-10	0.355751579708898\\
-5	0.219345653393144\\
0	0.126065383549774\\
5	0.0717676342032142\\
10	0.0368530079706634\\
15	0.0276470703960422\\
20	0.0200299084382422\\
};
\addlegendentry{Correlation-based, $L_h=3$}

\addplot [color=cyan, line width=2.0pt, mark=diamond, mark options={solid, cyan,mark size=4pt}]
  table[row sep=crcr]{%
-20	4.23331\\
-15	0.514686\\
-10	0.3583\\
-5	0.20596\\
0	0.1256\\
5	0.0707\\
10	0.0385\\
15	0.0254\\
20	0.0134\\
};
\addlegendentry{PIPIC \cite{marchese2024}}

\addplot [color=black, dashed, line width=2.0pt]
  table[row sep=crcr]{%
-20	1.29519441980234\\
-15	0.728341346410793\\
-10	0.409576437931569\\
-5	0.230321756873181\\
0	0.129519441980234\\
5	0.0728341346410793\\
10	0.0409576437931569\\
15	0.0230321756873181\\
20	0.0129519441980234\\
};
\addlegendentry{$\sqrt{\text{CRLB}(v)}$}

\end{axis}
\end{tikzpicture}%
}
    \caption{The range/velocity RMSE against radar SNR are shown for different values of $L_h$. The \gls{CRLB} is reported as a lower bound to radar parameter estimation.} 
    \label{fig:RMSE} 
\end{figure}

\subsection{Sensing Performance at the ISAC Transceiver}\label{SensingPerformance}  
To evaluate sensing performance, a scenario with a target at a range of $d=300$ m and a velocity of $v=70$ km/h is considered. The \gls{OTFS} frame parameters are set to $M=N=32$, and the transmitter employs \gls{QAM} signaling with $Q=4$.  

\subsubsection{Single-Target Range/Velocity RMSE vs Radar SNR}  
The radar parameter estimation performance of the proposed ISAC sensing algorithm is assessed using the \gls{RMSE} of the target's range and velocity.  

Fig.~\ref{fig:RMSE} presents the range and velocity RMSEs for different values of radar \gls{SNR}. The radar \gls{SNR} is defined as $\text{SNR}_{\text{rad}}=\frac{|g_i|^2}{N_0}$, where $|g_i|^2$ represents the target power. Performance is evaluated for different values of $L_h=1,2,3$ and compared against the \gls{CRLB} and the \gls{PIPIC} method \cite{marchese2024}, which performs joint \gls{DD} estimation.  
It can be observed that increasing $L_h$ allows the proposed approach to achieve performance close to the \gls{CRLB}. Specifically, sub-meter localization accuracy is attainable at $\text{SNR}_{\text{rad}}$ values above $10$ dB when $L_h\geq 2$. Moreover, the proposed method achieves performance comparable to that of the \gls{PIPIC}. Thus, the proposed disjoint \gls{DD} estimation preserves accuracy while lowering computational complexity.

\subsubsection{Target Discrimination in Multi-Target Scenarios}
As described in Section \ref{sec:multitarget}, the proposed \gls{FNN} architecture is also applicable to multi-target detection. In this context, Fig. \ref{fig:ResFNN} provides insights into \gls{DL}-based multi-target detection performance. Specifically, it can be observed that the proposed \gls{FNN} exhibits consistent detection capabilities regardless of the \gls{SNR}. Conversely, the performance of the \gls{SC}-based approach relies heavily on \gls{SNR} conditions. The proposed \gls{FNN} is capable of resolving targets that are at least one resolution cell apart across all \gls{SNR} levels. In contrast, the \gls{SC} fails to distinguish adjacent targets, incorrectly identifying them as a single entity in low \gls{SNR} scenarios even with larger delay separations.

\begin{figure}[t]
\centering
    \resizebox{0.99\columnwidth}{!}{
%
%
\definecolor{mycolor1}{rgb}{0.00000,0.44700,0.74100}%
\definecolor{mycolor2}{rgb}{0.85000,0.32500,0.09800}%
\definecolor{mycolor3}{rgb}{0.92900,0.69400,0.12500}%
\definecolor{mycolor4}{rgb}{0.49400,0.18400,0.55600}%
\definecolor{mycolor5}{rgb}{0.46600,0.67400,0.18800}%
\definecolor{mycolor6}{rgb}{0.30100,0.74500,0.93300}%
\begin{tikzpicture}

\begin{axis}[%
width=5in,
height=2.8565in,
at={(0.758in,0.482in)},
scale only axis,
xmin=0,
xmax=14,
xlabel style={font=\color{white!15!black}},
xlabel={\Large $E_b/N_0 \ \mathrm{[dB]}$},
ymode=log,
ymin=1e-4,
ymax=0.3,
yminorticks=true,
ylabel style={font=\color{white!15!black}},
ylabel={\Large $\mathrm{BER}$},
tick label style={font=\Large},
axis background/.style={fill=white},
axis x line=box,
axis y line=box,
xmajorgrids,
ymajorgrids,
yminorgrids,
legend style={at={(0.03,0.03)}, anchor=south west, legend cell align=left, align=left, draw=white!15!black, font=\large}
]


\addplot [color=mycolor3, line width=2pt, mark=o, mark options={solid, mycolor3, mark size=4pt}]
  table[row sep=crcr]{%
0	0.132367024739583\\
2	0.0926171875\\
4	0.0603722330729167\\
6	0.03637109375\\
8	0.0213108723958333\\
10	0.0129694010416667\\
12	0.00893424479166667\\
14	0.00731770833333333\\
};
\addlegendentry{LMMSE + TM  \cite{Raviteja2019,Viterbo2022}}

\addplot [color=mycolor4, line width=2pt, mark=triangle, mark options={solid, mycolor4, mark size=4pt}]
  table[row sep=crcr]{%
0	0.18430029296875\\
2	0.136817220052083\\
4	0.0955738932291667\\
6	0.0633289388020833\\
8	0.0423557942708333\\
10	0.0299161783854167\\
12	0.0234822591145833\\
14	0.0204638671875\\
};
\addlegendentry{IMFC + TM \cite{Raviteja2019,Viterbo2022}}

\addplot [color=mycolor5, line width=2pt, mark=o, mark options={solid, mycolor5, mark size=4pt}]
  table[row sep=crcr]{%
0	0.123682779947917\\
2	0.0820242513020833\\
4	0.0480558268229167\\
6	0.0234386393229167\\
8	0.00909326171875\\
10	0.00269938151041667\\
12	0.0005595703125\\
14	9.01692708333333e-05\\
};
\addlegendentry{LMMSE + CM-FNN}

\addplot [color=mycolor6, line width=2pt, mark=triangle, mark options={solid, mycolor6, mark size=4pt}]
  table[row sep=crcr]{%
0	0.158397298177083\\
2	0.105695475260417\\
4	0.06102197265625\\
6	0.0284259440104167\\
8	0.0101923828125\\
10	0.00267350260416667\\
12	0.000539876302083333\\
14	0.000111002604166667\\
};
\addlegendentry{IMFC + CM-FNN }

\addplot [color=mycolor1, dashed, line width=2pt]
  table[row sep=crcr]{%
0	0.123559733072917\\
2	0.08184765625\\
4	0.047880859375\\
6	0.0232555338541667\\
8	0.00898828125\\
10	0.00262434895833333\\
12	0.000543131510416667\\
14	8.31705729166667e-05\\
};
\addlegendentry{LMMSE, Perf. CSI}

\addplot [color=mycolor2, line width=2pt,dashed]
  table[row sep=crcr]{%
0	0.158798665364583\\
2	0.105798665364583\\
4	0.06085546875\\
6	0.0282644856770833\\
8	0.0101046549479167\\
10	0.00261474609375\\
12	0.000517903645833333\\
14	0.00010009765625\\
};
\addlegendentry{IMFC, Perf. CSI}

\end{axis}

\end{tikzpicture}%
}
    \caption{The bit error rate as a function of the signal-to-noise ratio per bit is shown for both the LMMSE and proposed IMFC equalizers in the following cases: perfect CSI at the receiver, imperfect CSI estimated through the baseline threshold method and through the proposed FNN-aided correlation-based method. The imperfect CSI curves serve as a sensitivity analysis to channel estimation errors.}
    \label{fig:BERvsSNR} 
\end{figure}

\subsection{Communication Performance}  
The receiver is assumed to execute a channel estimation algorithm to obtain a channel estimate, followed by an equalization method to compensate for the channel. Subsequently, a symbol-by-symbol \gls{ML} detector is employed to make hard decisions.  

\subsubsection{BER vs $E_b/N_0$}  
Communication performance is evaluated in terms of the \gls{BER} as a function of the signal-to-noise ratio per bit ($E_b/N_0$). Fig.~\ref{fig:BERvsSNR} illustrates the \gls{BER} as a function of $E_b/N_0$ for both the \gls{LMMSE} and the proposed \gls{IMFC} equalizers. In the simulations, the transmitter is assumed to use a $4$-\gls{QAM} constellation, while the \gls{IMFC} equalizer operates with $\epsilon=\frac{\sqrt{MN\sigma^2}}{2}$ and $n_{\max}=50$.  
The performance under perfect \gls{CSI} knowledge at the receiver is provided as a lower bound on the probability of error, serving as a reference for comparing the two equalization techniques. It can be observed that, at sufficiently high \gls{SNR}, the proposed \gls{IMFC} equalizer achieves a \gls{BER} comparable to that of the conventional \gls{LMMSE} equalizer.  
In order to investigate the sensitivity of the proposed \gls{IMFC} equalizer to channel estimation errors, Fig.~\ref{fig:BERvsSNR} also compares \gls{BER} performance under imperfect \gls{CSI}, where channel estimates are obtained from different estimation algorithms. Specifically, the \gls{BER} performance using \gls{LMMSE} and the proposed \gls{IMFC} equalizers is shown for channel estimates obtained via the conventional \gls{TM} method \cite{Raviteja2019,Viterbo2022} and the proposed \gls{CM-FNN}, assuming $\text{SNR}_p=18$ dB.  
It can be noted that the proposed channel estimation method in Algorithm \ref{alg:CM} achieves performance nearly equivalent to the ideal case of perfect \gls{CSI} for both the \gls{LMMSE} and \gls{IMFC} equalizers. In contrast, the conventional \gls{TM} method is constrained by \gls{IPI} and, therefore, fails to compensate for channel impairments when \gls{IPI} is significant. Consequently, the proposed \gls{CM-FNN} channel estimation algorithm achieves a substantial performance gain over the conventional \gls{TM} approach.  

\begin{figure}[t]
\centering
    \resizebox{0.99\columnwidth}{!}{
%
%
\definecolor{mycolor1}{rgb}{0.00000,0.44700,0.74100}%
\definecolor{mycolor2}{rgb}{0.85000,0.32500,0.09800}%
\definecolor{mycolor3}{rgb}{0.92900,0.69400,0.12500}%
\begin{tikzpicture}

\begin{axis}[%
width=5in,
height=2.5in,
at={(0.758in,0.482in)},
xmin=0,
xmax=14,
xlabel style={font=\color{white!15!black}},
xlabel={\Large $E_b/N_0 \ \mathrm{[dB]}$},
ymin=18,
ymax=32,
ylabel style={font=\color{white!15!black}},
ylabel={\Large Iterations},
tick label style={font=\Large},
axis background/.style={fill=white},
axis x line=box,
axis y line=box,
xmajorgrids,
ymajorgrids,
legend style={at={(1, 0.53)},anchor=east,legend cell align=left,draw=white!15!black,font=\large}
]
\addplot [color=mycolor1, line width=2pt, mark=triangle, mark options={solid, mycolor1, mark size=5pt}]
  table[row sep=crcr]{%
0	19.7336666666667\\
2	20.1306666666667\\
4	20.5026666666667\\
6	20.8753333333333\\
8	21.2436666666667\\
10	21.6073333333333\\
12	21.9243333333333\\
14	22.3023333333333\\
};
\addlegendentry{Perfect CSI}

\addplot [color=mycolor2, line width=2pt, mark=o, mark options={solid, mycolor2,mark size=4pt}]
  table[row sep=crcr]{%
0	27.7833333333333\\
2	28.0896666666667\\
4	28.3923333333333\\
6	28.703\\
8	29.031\\
10	29.346\\
12	29.798\\
14	30.6383333333333\\
};
\addlegendentry{Imperfect CSI, TM}

\addplot [color=mycolor3, line width=2pt, mark=x, mark options={solid, mycolor3,mark size=4pt}]
  table[row sep=crcr]{%
0	19.721\\
2	20.124\\
4	20.4923333333333\\
6	20.8666666666667\\
8	21.24\\
10	21.6026666666667\\
12	21.9343333333333\\
14	22.2883333333333\\
};
\addlegendentry{Imperfect CSI, CM-FNN}

\end{axis}

\end{tikzpicture}%
}
    \caption{The average number of iterations for the proposed IMFC equalizer as a function of the signal-to-noise ratio per bit. Both perfect CSI and imperfect CSI cases are considered using the baseline threshold method and the proposed FNN-aided correlation method.} 
    \label{fig:NavgvsSNR} 
\end{figure}

\begin{figure}[t]
\centering
    \resizebox{0.99\columnwidth}{!}{
        \begin{tikzpicture}
\definecolor{mycolor1}{rgb}{0.00000,0.44700,0.74100}%
\definecolor{mycolor2}{rgb}{0.85000,0.32500,0.09800}%

\begin{axis}[
    width=5.833in,
    height=2.49in,
    xmin=0.2,
    xmax=3,
    xlabel={\Large $\epsilon/\sqrt{MN\sigma^2}$},
    ymode=log,
    ymin=0.0001,
    ymax=0.01,
    ylabel={\Large BER},
    ylabel near ticks,
    xmajorgrids,
    ymajorgrids,
axis x line=box,
axis y line=box,
    tick label style={font=\Large},
    axis background/.style={fill=white}
]

\addplot[color=mycolor1, line width=2pt, mark=o, mark options={solid, mycolor1, mark size=4pt}]
    table[row sep=crcr]{%
0.2    0.000705729166666667\\
0.4    0.00072265625\\
0.6    0.000755696614583333\\
0.8    0.000837076822916667\\
1    0.000944498697916667\\
1.2    0.00107210286458333\\
1.4    0.00122737630208333\\
1.6    0.00146956380208333\\
1.8    0.00175113932291667\\
2    0.00212516276041667\\
2.2    0.002640625\\
2.4    0.00320621744791667\\
2.6    0.00388997395833333\\
2.8    0.00491031901041667\\
3    0.00621956380208333\\
};
\end{axis}

\begin{axis}[
    width=5.833in,
    height=2.49in,
    at={(0in,0in)},
    xmin=0.2,
    xmax=3,
    ymin=20,
    ymax=35,
    ylabel={\Large Iterations},
    ylabel near ticks,
    xmajorgrids,
    ymajorgrids,
    axis y line*=right,
    axis x line=none,
    tick label style={font=\large},
    axis background/.style={fill=none}
]
\addplot[color=mycolor2, line width=2pt, mark=o, mark options={solid, mycolor2, mark size=4pt}]
    table[row sep=crcr]{%
0.2    33.3316666666667\\
0.4    24.131\\
0.6    23.4906666666667\\
0.8    23.0026666666667\\
1    22.661\\
1.2    22.3876666666667\\
1.4    22.123\\
1.6    21.9093333333333\\
1.8    21.7163333333333\\
2    21.5116666666667\\
2.2    21.3423333333333\\
2.4    21.1606666666667\\
2.6    20.992\\
2.8    20.8376666666667\\
3    20.6813333333333\\
};
\end{axis}

\node[
    anchor=north east,
    font=\Large,
    xshift=5.15in,
    yshift=1in,
    draw,
    fill=white,
    inner sep=0pt
] (legend) {
    \begin{tabular}{ll}
        \tikz{\draw[mycolor1, line width=2pt] (0,0) -- (0.5,0); \draw[mycolor1, line width=2pt] (0.25,0) circle (4pt);}  BER \\
        \tikz{\draw[mycolor2, line width=2pt] (0,0) -- (0.5,0); \draw[mycolor2, line width=2pt] (0.25,0) circle (4pt);}  Iterations \\
    \end{tabular}
};

\end{tikzpicture}}
    \caption{The influence of the threshold $\epsilon$ on the average number of iterations and BER for the proposed IMFC equalizer.} 
    \label{fig:ThInfluence} 
\end{figure}

\subsubsection{Average Number of Iterations for the IMFC Equalizer}  
To assess the complexity of the proposed equalization scheme compared to the \gls{LMMSE}, the average number of iterations is analyzed as a function of $E_b/N_0$. Fig.~\ref{fig:NavgvsSNR} presents the average number of iterations required by the proposed \gls{IMFC} equalizer under three conditions: perfect \gls{CSI} at the receiver, and imperfect \gls{CSI} estimated using the conventional \gls{TM} and the proposed \gls{CM-FNN}.  
It can be observed that the complexity of the \gls{IMFC} equalizer remains unaffected by imperfect \gls{CSI} when using the proposed channel estimation algorithm. Conversely, complexity increases when \gls{CSI} is estimated using the conventional \gls{TM} method. However, it is worth noting that the average number of iterations required by the \gls{IMFC} equalizer remains significantly lower than $MN^2/P$.  
Therefore, as discussed in Section \ref{sec:complexityIMFC}, the proposed \gls{IMFC} equalizer offers substantially lower complexity than the conventional \gls{LMMSE} while achieving nearly the same \gls{BER} performance.

\subsubsection{Influence of the Threshold}  
The choice of the threshold $\epsilon$ for the stopping criterion of the \gls{IMFC} equalizer plays a crucial role in balancing the performance/complexity trade-off. Specifically, a higher threshold is expected to reduce complexity at the cost of degraded \gls{BER} performance, whereas a lower threshold may enhance \gls{BER} performance but at the expense of increased complexity.  
Fig.~\ref{fig:ThInfluence} illustrates the average number of iterations and \gls{BER} performance of the proposed \gls{IMFC} equalizer for different values of the threshold $\epsilon$. Simulations are conducted at $E_b/N_0=12$ dB. It can be observed that \gls{BER} decreases as $\epsilon$ is reduced. However, the improvement stabilizes around $\epsilon/\sqrt{MN\sigma^2}=0.5$. Conversely, complexity starts to increase significantly when the threshold is further reduced.  
Therefore, the optimal threshold choice is $\epsilon=0.5\sqrt{MN\sigma^2}$. This analysis justifies the selected threshold value used for evaluating \gls{BER} performance.

\section{Conclusions}  
In this work, the \gls{DD} input-output relationship for  \gls{OTFS}-\gls{ISAC} with \gls{RCP} is formulated by establishing an exact algebraic decoupling of fractional channel parameters. This structural separation enables a low-complexity, disjoint \gls{DD} estimation that operates completely free from any asymptotic approximations, or restrictive assumptions on the transmitted \gls{DD} signal. The \gls{DD} channel effects are represented through a unitary sparse matrix, enabling simple correlation-based channel estimation using matched filtering. To further enhance performance, a \gls{FNN}-based architecture is introduced to estimate the number of propagation paths, overcoming the limitations of conventional stopping criteria and enhancing performance under low \gls{SNR} conditions.   
Simulation results demonstrate that the proposed \gls{FNN}-aided correlation-based approach achieves accurate channel estimation and \gls{BER} performance comparable to the ideal case with perfect \gls{CSI} knowledge.  
Sensing performance is evaluated by computing the \gls{RMSE}, and the results show that increasing the number of hierarchical levels allows performance close to the \gls{CRLB} to be achieved. Therefore, disjoint estimation allows reducing complexity without degrading performance, compared to joint estimation algorithms.  
Furthermore, by exploiting the sparsity of the \gls{DD} parameter matrix, a reduced-complexity equalizer based on Landweber method, called \gls{IMFC} algorithm, is proposed for channel equalization. This approach achieves \gls{BER} performance comparable to that of the conventional \gls{LMMSE} equalizer but with significantly lower complexity. 

Future works will consider disjoint fractional \gls{DD} estimation in multi-antenna \gls{OTFS} receivers using superimposed pilot schemes for \gls{PAPR} reduction and increased spectral efficiency against exclusive pilot schemes.

\section*{Appendix A}
\section*{Proof of theorem \ref{Th1}}
The cyclic shift property in \eqref{eq:RxMultipath} inherited by the \gls{RCP} allows a frequency-domain representation of the transmitted signal \cite{Keskin2024,Keskin2021}, thus $s([t-\tau_i]_{NT})=\mathcal{F}^{-1}\{S(f)e^{-j2\pi f\tau_i}\}$. Sampling \eqref{eq:RxMultipath} at $T/M$ and discretizing frequency (with spacing $\Delta f/N$), the following samples are obtained
\begin{equation} \label{eq:RxHadamard}
\mathbf{r}=\sum_{i=1}^Pg_i[\mathbf{F}_{MN}^{\mathsf{H}}(\mathbf{F}_{MN}\mathbf{s}\odot\mathbf{b}(\tau_i))]\odot\mathbf{c}(\nu_i)+\mathbf{n},
\end{equation}
where $\mathbf{b}(\tau_i)=[e^{-j2\pi q\frac{\Delta f}{N}\tau_i}]_{q=0}^{MN-1}$ and $\mathbf{c}(\nu_i)=[e^{j2\pi q\frac{T}{M}\nu_i}]_{q=0}^{MN-1}$ are the frequency-domain and temporal steering vectors, respectively. By normalizing channel parameters as
$l_i=\frac{\tau_i}{\Delta\tau}=L_i+\tilde{l}_i, \ k_i=\frac{\nu_i}{\Delta\nu}=K_i+\tilde{k}_i$, the expression for the elements of $\mathbf{d}(\tau_i)$ and $\mathbf{c}(\nu_i)$, it is possible to notice that 
\begin{equation} \label{eq:Exponentials1}
e^{-j2\pi q\frac{\Delta f}{N}\tau_i}=e^{-j2\pi q\frac{\Delta f}{N}(L_i+\tilde{l}_i))\Delta\tau}=\Big(e^{-j\frac{2\pi}{MN}q}\Big)^{L_i+\tilde{l}_i},
\end{equation}
\begin{equation} \label{eq:Exponentials2}
e^{j2\pi q\frac{T}{M}\nu_i}=e^{j2\pi q\frac{T}{M}(K_i+\tilde{k}_i))\Delta\nu}=\Big(e^{j\frac{2\pi}{MN}q}\Big)^{K_i+\tilde{k}_i}.
\end{equation}
Therefore, the two Hadamard products in \eqref{eq:RxHadamard} can be expressed as a matrix by vector product by defining the diagonal matrix $\mathbf{D}=\textnormal{diag}\Big\{\Big[e^{j\frac{2\pi}{MN}q}\Big]_{q=0}^{MN-1}\Big\}$ as
\begin{equation} \label{eq:RxHadamardDMtx}
\mathbf{r}=\sum_{i=1}^Pg_i\mathbf{D}^{K_i}\mathbf{D}^{\tilde{k}_i}\mathbf{F}_{MN}^{\mathsf{H}}(\mathbf{D}^*)^{L_i}(\mathbf{D}^*)^{\tilde{l}_i}\mathbf{F}_{MN}\mathbf{s}+\mathbf{n}.
\end{equation}
Finally, \eqref{eq:DelayTimeChannelMtx} is obtained from \eqref{eq:DelayTimeIO} and \eqref{eq:RxHadamardDMtx}.

\section*{Appendix B}
\section*{Proof of Corollary \ref{Corol1}}
If delays and Dopplers are multiple of delay and Doppler resolutions, it can be noticed that $\mathbf{D}^{\tilde{k}_i}=\mathbf{I}_{MN}$ and $(\mathbf{D}^*)^{\tilde{l}_i}=\mathbf{I}_{MN}$. Hence, \eqref{eq:DelayTimeChannelMtx} reduces to
\begin{equation} \label{eq:DelayTimeChannelMtxIntDDsEigen}
\mathbf{H}=\sum_{i=1}^Pg_i\mathbf{D}^{K_i}\mathbf{F}_{MN}^{\mathsf{H}}(\mathbf{D}^*)^{L_i}\mathbf{F}_{MN}.
\end{equation}
Moreover, it can be noted that $\mathbf{F}_{MN}^{\mathsf{H}}(\mathbf{D}^*)^{L_i}\mathbf{F}_{MN}$ is the eigenvalue decomposition of the circulant matrix $\mathbf{\Pi}^{L_i}$. In fact, $\mathbf{\Pi}^{L_i}=\mathbf{F}_{MN}^{\mathsf{H}}\mathbf{\Lambda}\mathbf{F}_{MN}$ where the diagonal matrix $\mathbf{\Lambda}$ is given as
\begin{equation}\begin{split}\label{eq:EigenDiagonal}
\mathbf{\Lambda}&=\text{diag}\Big(\sqrt{MN}\mathbf{F}_{MN}\mathbf{e}_{[L_i+1]_{MN}}\Big)
\\
&=\text{diag}\Big(\big[z^{-qL_i}\big]_{q=0}^{MN-1}\Big)=(\mathbf{D}^*)^{L_i}.
\end{split}
\end{equation}

\section*{Appendix C}
\section*{Proof of Proposition \ref{lem:Qexpression}}
\label{App:QProof}
The generic element of $\mathbf{Q}(a)$ is obtained by 
\begin{equation}\label{eq:elementQ_matrixprod}
[\mathbf{Q}(a)]_{p,q} = \sum_{s=0}^{MN-1} [\tilde{\mathbf{F}}]_{p,s} \sum_{t=0}^{MN-1}[\tilde{\mathbf{D}}(a)]_{s,t} [\tilde{\mathbf{F}}]_{t,q},
\end{equation}
where $\tilde{\mathbf{F}}=\mathbf{F}_{N}\otimes\mathbf{I}_{M}$, $\tilde{\mathbf{D}}(a) = \mathbf{D}^a\mathbf{F}_{MN}^{\mathsf{H}}$ and
\begin{equation}\begin{split}\label{eq:Ftilde}
[\tilde{\mathbf{F}}]_{t,q} &= [\mathbf{F}_N]_{\lfloor t/M \rfloor, \lfloor q/M \rfloor} [\mathbf{I}_{M}]_{\text{mod}(t,M),\text{mod}(q,M)} \\
&=[\mathbf{F}_N]_{\lfloor t/M \rfloor, \lfloor q/M \rfloor} \delta_{\text{mod}(t,M),\text{mod}(q,M)},
\end{split}\end{equation}
\begin{equation}\begin{split}\label{eq:Dtilde}
[\tilde{\mathbf{D}}(a)]_{s,t} &= \\ \sum_{r=0}^{MN-1} &[\mathbf{D}^a]_{s,r} [\mathbf{F}_{MN}^{\mathsf{H}}]_{r,t}= [\mathbf{D}^a]_{s,s} [\mathbf{F}_{MN}^{\mathsf{H}}]_{s,t}=\frac{e^{j\frac{2\pi s(a+t)}{MN}}}{\sqrt{MN}}.
\end{split}\end{equation}
Plugging \eqref{eq:Dtilde} and \eqref{eq:Ftilde} back in \eqref{eq:elementQ_matrixprod} 
\begin{equation}\begin{split}\label{eq:elementQ_matrixprod_substitution}
[\mathbf{Q}(a)]_{p,q}= &\\
 \frac{1}{N\sqrt{MN}}& \sum_{s=0}^{MN-1}[\mathbf{F}_N]_{\lfloor p/M \rfloor, \lfloor s/M \rfloor} \delta_{\text{mod}(p,M),\text{mod}(s,M)} \\
\sum_{t=0}^{MN-1} & e^{j\frac{2\pi s(a+t)}{MN}} [\mathbf{F}_N]_{\lfloor t/M \rfloor, \lfloor q/M \rfloor} \delta_{\text{mod}(t,M),\text{mod}(q,M)}.
\end{split}\end{equation}
By introducing new indices $k_1$ and $k_2$ ranging from $0$ to $N-1$ and substituting $s$ and $t$ as $s = k_1 M + \text{mod}(p,M)$ and $t = k_2 M + \text{mod}(q,M)$ in \eqref{eq:elementQ_matrixprod_substitution}, the final form presented in \eqref{eq:Qelem_lemma} in Proposition \ref{lem:Qexpression} is obtained.
Moreover, \eqref{eq:Qelem_lemma} can be rewritten as
\begin{equation}\begin{split}\label{eq:Qelem_sparsity}
\big[\mathbf{Q}(a)\big]_{p,q} &= \frac{1}{N\sqrt{MN}} \\
\sum_{k_1=0}^{N-1} & e^{-j\frac{2\pi \lfloor p/M \rfloor k_1}{N}}e^{j\frac{2\pi}{MN}\big(k_1M+\text{mod}(p,M)\big)\big(a+\text{mod}(q,M)\big)} 
\\  \sum_{k_2=0}^{N-1} & e^{j\frac{2\pi k_2}{N}\big(k_1M+\text{mod}(p,M)-\lfloor q/M \rfloor\big)}.
\end{split}
\end{equation}
Finally, using the fact that $\sum_{k'=0}^{N-1} e^{j\frac{2\pi k' u}{N}}=N$ if $u$ is an integer multiple of $N$ and zero otherwise, it can be shown that, if $M$ is an integer multiple of $N$, the element $\big[\mathbf{Q}(a)\big]_{p,q}$ is non-zero, for a fixed row $p$, for $M$ consecutive values of $q$.

\ifCLASSOPTIONcaptionsoff
  \newpage
\fi

\balance
\bibliographystyle{IEEEtran}
\bibliography{list_IEEE}

@ARTICLE{Muppaneni2023,
  author    = {Muppaneni, Sai Pradeep and Mattu, Sandesh Rao and Chockalingam, A.},
  journal   = {IEEE Commun. Lett.},
  title     = {{Channel and Radar Parameter Estimation With Fractional Delay-Doppler Using {OTFS}}},
  year      = {2023},
  volume    = {27},
  number    = {5},
  pages     = {1392--1396},
  doi       = {10.1109/LCOMM.2023.3251578}
}

@ARTICLE{Raviteja2019,
  author    = {Raviteja, P. and Phan, Khoa T. and Hong, Yi},
  journal   = {IEEE Trans. Veh. Technol.},
  title     = {{Embedded Pilot-Aided Channel Estimation for {OTFS} in Delay--Doppler Channels}},
  year      = {2019},
  volume    = {68},
  number    = {5},
  pages     = {4906--4917},
  doi       = {10.1109/TVT.2019.2906357}
}

@INPROCEEDINGS{Hadani2017,
  author    = {Hadani, R. and Rakib, S. and Tsatsanis, M. and Monk, A. and Goldsmith, A. J. and Molisch, A. F. and Calderbank, R.},
  booktitle = {2017 IEEE Wireless Commun. Netw. Conf. (WCNC)},
  title     = {{Orthogonal Time Frequency Space Modulation}},
  year      = {2017},
  pages     = {1--6},
  doi       = {10.1109/WCNC.2017.7925924}
}

@misc{Khan2021,
  author       = {Khan, Imran Ali and Mohammed, Saif Khan},
  title        = {{Low Complexity Channel Estimation for {OTFS} Modulation with Fractional Delay and Doppler}},
  howpublished = {arXiv:2111.06009 [cs.IT]},
  year         = {2021}
}

@ARTICLE{Yogesh2024,
  author    = {Yogesh, Vineetha and Mattu, Sandesh Rao and Chockalingam, A.},
  journal   = {IEEE Commun. Lett.},
  title     = {{Low-Complexity Delay-Doppler Channel Estimation in Discrete {Zak} Transform Based {OTFS}}},
  year      = {2024},
  volume    = {28},
  number    = {3},
  pages     = {672--676},
  doi       = {10.1109/LCOMM.2024.3351685}
}

@ARTICLE{Mattu2024,
  author    = {Mattu, Sandesh Rao and Chockalingam, A.},
  journal   = {IEEE Wireless Commun. Lett.},
  title     = {{Learning in Time-Frequency Domain for Fractional Delay-Doppler Channel Estimation in {OTFS}}},
  year      = {2024},
  volume    = {13},
  number    = {5},
  pages     = {1245--1249},
  doi       = {10.1109/LWC.2024.3367112}
}

@book{Viterbo2022,
  author    = {Hong, Yi and Thaj, Tharaj and Viterbo, Emanuele},
  title     = {{Delay-Doppler Communications: Principles and Applications}},
  publisher = {Elsevier},
  address   = {Netherlands},
  year      = {2022},
  isbn      = {9780323859660},
  doi       = {10.1016/C2020-0-01791-3}
}

@ARTICLE{Giordani2020,
  author    = {Giordani, Marco and Polese, Michele and Mezzavilla, Marco and Rangan, Sundeep and Zorzi, Michele},
  journal   = {IEEE Commun. Mag.},
  title     = {{Toward {6G} Networks: Use Cases and Technologies}},
  year      = {2020},
  volume    = {58},
  number    = {3},
  pages     = {55--61},
  doi       = {10.1109/MCOM.001.1900411}
}

@ARTICLE{Zhang2019,
  author    = {Zhang, Zhengquan and Xiao, Yue and Ma, Zheng and Xiao, Ming and Ding, Zhiguo and Lei, Xianfu and Karagiannidis, George K. and Fan, Pingzhi},
  journal   = {IEEE Veh. Technol. Mag.},
  title     = {{{6G} Wireless Networks: Vision, Requirements, Architecture, and Key Technologies}},
  year      = {2019},
  volume    = {14},
  number    = {3},
  pages     = {28--41},
  doi       = {10.1109/MVT.2019.2921208}
}

@ARTICLE{Wang2006,
  author    = {Wang, Tiejun and Proakis, J. G. and Masry, E. and Zeidler, J. R.},
  journal   = {IEEE Trans. Wireless Commun.},
  title     = {{Performance Degradation of {OFDM} Systems Due to Doppler Spreading}},
  year      = {2006},
  volume    = {5},
  number    = {6},
  pages     = {1422--1432},
  doi       = {10.1109/TWC.2006.1638663}
}

@ARTICLE{Raviteja2018,
  author    = {Raviteja, P. and Phan, Khoa T. and Hong, Yi and Viterbo, Emanuele},
  journal   = {IEEE Trans. Wireless Commun.},
  title     = {{Interference Cancellation and Iterative Detection for Orthogonal Time Frequency Space Modulation}},
  year      = {2018},
  volume    = {17},
  number    = {10},
  pages     = {6501--6515},
  doi       = {10.1109/TWC.2018.2860011}
}

@ARTICLE{Tiwiri2019,
  author    = {Tiwari, Shashank and Das, Suvra Sekhar and Rangamgari, Vivek},
  journal   = {IEEE Commun. Lett.},
  title     = {{Low Complexity {LMMSE} Receiver for {OTFS}}},
  year      = {2019},
  volume    = {23},
  number    = {12},
  pages     = {2205--2209},
  doi       = {10.1109/LCOMM.2019.2945564}
}

@ARTICLE{Surabhi2020,
  author    = {Surabhi, G. D. and Chockalingam, A.},
  journal   = {IEEE Commun. Lett.},
  title     = {{Low-Complexity Linear Equalization for {OTFS} Modulation}},
  year      = {2020},
  volume    = {24},
  number    = {2},
  pages     = {330--334},
  doi       = {10.1109/LCOMM.2019.2956709}
}

@ARTICLE{Zhao2020,
  author    = {Zhao, Lei and Gao, Wen-Jing and Guo, Wenbin},
  journal   = {IEEE Commun. Lett.},
  title     = {{Sparse Bayesian Learning of Delay-Doppler Channel for {OTFS} System}},
  year      = {2020},
  volume    = {24},
  number    = {12},
  pages     = {2766--2769},
  doi       = {10.1109/LCOMM.2020.3021120}
}

@ARTICLE{Wei2022,
  author    = {Wei, Zhiqiang and Yuan, Weijie and Li, Shuangyang and Yuan, Jinhong and Ng, Derrick Wing Kwan},
  journal   = {IEEE Trans. Wireless Commun.},
  title     = {{Off-Grid Channel Estimation With Sparse Bayesian Learning for {OTFS} Systems}},
  year      = {2022},
  volume    = {21},
  number    = {9},
  pages     = {7407--7426},
  doi       = {10.1109/TWC.2022.3158616}
}

@INPROCEEDINGS{Li2022,
  author    = {Li, Qingyu and Gong, Yi and Meng, Fanke and Li, Zhongjie and Miao, Linlong and Xu, Zhan},
  booktitle = {2022 IEEE/CIC Int. Conf. Commun. China (ICCC Workshops)},
  title     = {{Residual Learning Based Channel Estimation for {OTFS} System}},
  year      = {2022},
  pages     = {275--280},
  doi       = {10.1109/ICCCWorkshops55477.2022.9896637}
}

@ARTICLE{marchese2024,
  author    = {Marchese, M. and Wymeersch, H. and Spallaccini, P. and Savazzi, P.},
  journal   = {Sensors},
  title     = {{Progressive Inter-Path Interference Cancellation Algorithm for Channel Estimation Using Orthogonal Time--Frequency Space}},
  year      = {2024},
  volume    = {24},
  number    = {13},
  pages     = {4414},
  doi       = {10.3390/s24134414}
}

@ARTICLE{Gaudio2020,
  author    = {Gaudio, L. and Kobayashi, M. and Caire, G. and Colavolpe, G.},
  journal   = {IEEE Trans. Wireless Commun.},
  title     = {{On the Effectiveness of {OTFS} for Joint Radar Parameter Estimation and Communication}},
  year      = {2020},
  volume    = {19},
  number    = {9},
  month     = {sep},
  doi       = {10.1109/TWC.2020.2998583}
}

@INPROCEEDINGS{Naikoti2021_det,
  author    = {Naikoti, Ashwitha and Chockalingam, A.},
  booktitle = {2021 IEEE 93rd Veh. Technol. Conf. (VTC2021-Spring)},
  title     = {{Low-Complexity Delay-Doppler Symbol {DNN} for {OTFS} Signal Detection}},
  year      = {2021},
  pages     = {1--6},
  doi       = {10.1109/VTC2021-Spring51267.2021.9448630}
}

@INPROCEEDINGS{Mattu2022,
  author    = {Mattu, Sandesh Rao and Chockalingam, A.},
  booktitle = {2022 IEEE 96th Veh. Technol. Conf. (VTC2022-Fall)},
  title     = {{Learning Based Delay-Doppler Channel Estimation with Interleaved Pilots in {OTFS}}},
  year      = {2022},
  pages     = {1--6},
  doi       = {10.1109/VTC2022-Fall57202.2022.10012974}
}

@INPROCEEDINGS{Zhang2022_adth,
  author    = {Zhang, Xiaoqi and Yuan, Weijie and Liu, Chang and Liu, Fan and Wen, Miaowen},
  booktitle = {2022 Int. Symp. Wireless Commun. Syst. (ISWCS)},
  title     = {{Deep Learning with a Self-Adaptive Threshold for {OTFS} Channel Estimation}},
  year      = {2022},
  pages     = {1--5},
  doi       = {10.1109/ISWCS56560.2022.9940260}
}

@INPROCEEDINGS{Zhang2022_AN,
  author    = {Zhang, Xiaoqi and Yuan, Weijie and Liu, Chang},
  booktitle = {2022 IEEE/CIC Int. Conf. Commun. China (ICCC Workshops)},
  title     = {{Deep Residual Learning for {OTFS} Channel Estimation with Arbitrary Noise}},
  year      = {2022},
  pages     = {320--324},
  doi       = {10.1109/ICCCWorkshops55477.2022.9896721}
}

@INPROCEEDINGS{Gao2024,
  author    = {Gao, Yuan and Xu, Xinchen and Hu, Bintao and Du, Jianbo and Yang, Wenrui and Jin, Yanliang},
  booktitle = {2024 33rd Int. Conf. Comput. Commun. Netw. (ICCCN)},
  title     = {{Channel Estimation for {MIMO}-{OTFS} Satellite Communication: A Deep Learning-Based Approach}},
  year      = {2024},
  pages     = {1--6},
  doi       = {10.1109/ICCCN61486.2024.10637519}
}

@ARTICLE{Raviteja2019_2,
  author    = {Raviteja, P. and Hong, Yi and Viterbo, Emanuele and Biglieri, Ezio},
  journal   = {IEEE Trans. Veh. Technol.},
  title     = {{Practical Pulse-Shaping Waveforms for Reduced-Cyclic-Prefix {OTFS}}},
  year      = {2019},
  volume    = {68},
  number    = {1},
  pages     = {957--961},
  doi       = {10.1109/TVT.2018.2878891}
}

@ARTICLE{Keskin2024,
  author    = {Keskin, Musa Furkan and Marcus, Carina and Eriksson, Olof and Alvarado, Alex and Widmer, Joerg and Wymeersch, Henk},
  journal   = {IEEE Trans. Wireless Commun.},
  title     = {{Integrated Sensing and Communications With {MIMO}-{OTFS}: {ISI}/{ICI} Exploitation and Delay-{Doppler} Multiplexing}},
  year      = {2024},
  volume    = {23},
  number    = {8},
  pages     = {10229--10246},
  doi       = {10.1109/TWC.2024.3370501}
}

@ARTICLE{Khan2023,
  author    = {Khan, Imran Ali and Mohammed, Saif Khan},
  journal   = {IEEE Wireless Commun. Lett.},
  title     = {{A Low-Complexity {OTFS} Channel Estimation Method for Fractional Delay-Doppler Scenarios}},
  year      = {2023},
  volume    = {12},
  number    = {9},
  pages     = {1484--1488},
  doi       = {10.1109/LWC.2023.3274936}
}

@INPROCEEDINGS{Pfadler2021,
  author    = {Pfadler, Andreas and Jung, Peter and Szollmann, Tom and Stanczak, Slawomir},
  booktitle = {2021 IEEE Int. Conf. Commun. Workshops (ICC Workshops)},
  title     = {{Pulse-Shaped {OTFS} Over Doubly-Dispersive Channels: One-Tap vs. Full {LMMSE} Equalizers}},
  year      = {2021},
  pages     = {1--6},
  doi       = {10.1109/ICCWorkshops50388.2021.9473535}
}

@ARTICLE{Wei2021,
  author    = {Wei, Zhiqiang and Yuan, Weijie and Li, Shuangyang and Yuan, Jinhong and Ng, Derrick Wing Kwan},
  journal   = {IEEE Trans. Commun.},
  title     = {{Transmitter and Receiver Window Designs for Orthogonal Time-Frequency Space Modulation}},
  year      = {2021},
  volume    = {69},
  number    = {4},
  pages     = {2207--2223},
  doi       = {10.1109/TCOMM.2021.3051386}
}

@INPROCEEDINGS{Rasheed2022,
  author    = {Rasheed, O. K. and Surabhi, G. D. and Chockalingam, A.},
  booktitle = {2020 IEEE 91st Veh. Technol. Conf. (VTC2020-Spring)},
  title     = {{Sparse Delay-Doppler Channel Estimation in Rapidly Time-Varying Channels for Multiuser {OTFS} on the Uplink}},
  year      = {2020},
  pages     = {1--5},
  doi       = {10.1109/VTC2020-Spring48590.2020.9128497}
}

@ARTICLE{Guo2023,
  author    = {Guo, Lin and Gu, Peng and Zou, Jun and Liu, Guangzu and Shu, Feng},
  journal   = {IEEE Trans. Veh. Technol.},
  title     = {{{DNN}-Based Fractional Doppler Channel Estimation for {OTFS} Modulation}},
  year      = {2023},
  volume    = {72},
  number    = {11},
  pages     = {15062--15067},
  doi       = {10.1109/TVT.2023.3280901}
}

@ARTICLE{Zhang2024sbl,
  author    = {Zhang, Yang and Zhang, Qunfei and He, Chengbing and Jing, Lianyou and Zheng, Tonghui and Yuen, Chau},
  journal   = {IEEE Trans. Veh. Technol.},
  title     = {{Sparse Bayesian Learning Approach for {OTFS} Channel Estimation With Fractional Doppler}},
  year      = {2024},
  volume    = {73},
  number    = {11},
  pages     = {16846--16860},
  doi       = {10.1109/TVT.2024.3420136}
}

@ARTICLE{Mohammed2022,
  author    = {Mohammed, Saif Khan and Hadani, Ronny and Chockalingam, Ananthanarayanan and Calderbank, Robert},
  journal   = {IEEE BITS Inf. Theory Mag.},
  title     = {{{OTFS}---A Mathematical Foundation for Communication and Radar Sensing in the Delay-Doppler Domain}},
  year      = {2022},
  volume    = {2},
  number    = {2},
  pages     = {36--55},
  doi       = {10.1109/MBITS.2022.3216536}
}

@misc{HadaniMonk2018,
  author       = {Hadani, Ronny and Monk, Anton},
  title        = {{{OTFS}: A New Generation of Modulation Addressing the Challenges of {5G}}},
  howpublished = {arXiv:1802.02623 [cs.IT]},
  year         = {2018}
}

@ARTICLE{Wei2022papr,
  author    = {Wei, Peng and Xiao, Yue and Feng, Wei and Ge, Ning and Xiao, Ming},
  journal   = {IEEE Trans. Wireless Commun.},
  title     = {{Characterizing the Peak-to-Average Power Ratio of {OTFS} Signals: A Large System Analysis}},
  year      = {2022},
  volume    = {21},
  number    = {6},
  pages     = {3705--3720},
  doi       = {10.1109/TWC.2021.3123397}
}

@INPROCEEDINGS{raviteja2019Radar,
  author    = {Raviteja, P. and Phan, Khoa T. and Hong, Yi and Viterbo, Emanuele},
  booktitle = {2019 IEEE Radar Conf. (RadarConf)},
  title     = {{Orthogonal Time Frequency Space ({OTFS}) Modulation Based Radar System}},
  year      = {2019},
  pages     = {1--6},
  doi       = {10.1109/RADAR.2019.8835764}
}

@INPROCEEDINGS{Keskin2021,
  author    = {Keskin, Musa Furkan and Wymeersch, Henk and Alvarado, Alex},
  booktitle = {2021 IEEE Int. Conf. Commun. Workshops (ICC Workshops)},
  title     = {{Radar Sensing with {OTFS}: Embracing {ISI} and {ICI} to Surpass the Ambiguity Barrier}},
  year      = {2021},
  pages     = {1--6},
  doi       = {10.1109/ICCWorkshops50388.2021.9473534}
}

@INPROCEEDINGS{Gaudio2020RadarMIMO,
  author    = {Gaudio, Lorenzo and Kobayashi, Mari and Caire, Giuseppe and Colavolpe, Giulio},
  booktitle = {2020 IEEE Radar Conf. (RadarConf20)},
  title     = {{Joint Radar Target Detection and Parameter Estimation with {MIMO} {OTFS}}},
  year      = {2020},
  pages     = {1--6},
  doi       = {10.1109/RadarConf2043947.2020.9266546}
}

@INPROCEEDINGS{Dehkordi2022,
  author    = {Dehkordi, Saeid K. and Gaudio, Lorenzo and Kobayashi, Mari and Colavolpe, Giulio and Caire, Giuseppe},
  booktitle = {2022 IEEE Int. Conf. Commun. Workshops (ICC Workshops)},
  title     = {{Beam-Space {MIMO} Radar with {OTFS} Modulation for Integrated Sensing and Communications}},
  year      = {2022},
  pages     = {509--514},
  doi       = {10.1109/ICCWorkshops53468.2022.9814573}
}

@INPROCEEDINGS{marchese2025reducedlat,
  author    = {Marchese, Mauro and Wymeersch, Henk and Spallaccini, Paolo and Chinnici, Stefano and Savazzi, Pietro},
  booktitle = {2025 IEEE Int. Conf. Mach. Learn. Commun. Netw. (ICMLCN)},
  title     = {{Reduced-Latency DL-Based Fractional Channel Estimation in {OTFS} Receivers}},
  year      = {2025},
  pages     = {1--6},
  doi       = {10.1109/ICMLCN64995.2025.11140575}
}

@ARTICLE{Li2025,
  author    = {Li, Xiangjun and Liang, Yu and Zhou, Zhengchun and Fan, Pingzhi},
  journal   = {IEEE Commun. Lett.},
  title     = {{Fractional Delay-Doppler Channel Estimation for {OTFS} Systems Based on Segmentation Technique and Sparse Bayesian Learning}},
  year      = {2025},
  volume    = {29},
  number    = {5},
  pages     = {1067--1071},
  doi       = {10.1109/LCOMM.2025.3553831}
}

@ARTICLE{serrano2025,
  author    = {Correas-Serrano, Aitor and Petrov, Nikita and Gonzalez-Huici, Maria and Yarovoy, Alexander},
  journal   = {IEEE Aerosp. Electron. Syst. Mag.},
  title     = {{Emerging Trends in Radar: {OTFS}-Based Radar for Integrated Sensing and Communications Systems}},
  year      = {2025},
  volume    = {40},
  number    = {6},
  pages     = {102--107},
  doi       = {10.1109/MAES.2025.3537576}
}

@ARTICLE{singh2025,
  author    = {Singh, Shivani and Nakkeeran, Amudheesan and Singh, Prem and Sharma, Ekant and Bapat, Jyotsna},
  journal   = {IEEE Trans. Veh. Technol.},
  title     = {{Target Detection for {OTFS}-Aided Cell-Free {MIMO} {ISAC} System}},
  year      = {2025}, 
  volume    = {74},
  number    = {7},
  pages     = {11568--11573},
  doi       = {10.1109/TVT.2025.3550135}
}

@ARTICLE{song2025,
  author    = {Song, Guangbo and Bai, Jiahao and Wang, Xinyi and Wei, Guohua and Yuan, Weijie and Quek, Tony Q. S.},
  journal   = {IEEE Trans. Commun.},
  title     = {{Low Sidelobe Level and {PAPR} {OTFS} Waveform Design for {ISAC} Systems}},
  year      = {2025},
  volume    = {73},
  number    = {9},
  pages     = {7952--7966},
  doi       = {10.1109/TCOMM.2025.3548760}
}

@ARTICLE{Almahdawi2022,
  author    = {Al-Mahdawi, H. K. I. and Alkattan, H. and Abotaleb, M. and Kadi, A. and El-kenawy, E.-S. M.},
  journal   = {Mathematics},
  title     = {{Updating the Landweber Iteration Method for Solving Inverse Problems}},
  year      = {2022},
  volume    = {10},
  number    = {15},
  pages     = {2798},
  doi       = {10.3390/math10152798}
}

@ARTICLE{CharlesByrne_2004,
  author    = {Byrne, Charles},
  journal   = {Inverse Problems},
  title     = {{A Unified Treatment of Some Iterative Algorithms in Signal Processing and Image Reconstruction}},
  year      = {2003},
  month     = {nov},
  volume    = {20},
  number    = {1},
  pages     = {103--120},
  doi       = {10.1088/0266-5611/20/1/006}
}

@ARTICLE{Chong2025,
  author    = {Chong, Ruoxi and Li, Shuangyang and Wei, Zhiqiang and Matthaiou, Michail and Ng, Derrick Wing Kwan and Caire, Giuseppe},
  journal   = {IEEE Trans. Commun.},
  title     = {{Cross-Domain Iterative Detection for {OTFS} Transmission With Frequency Domain Equalization}},
  year      = {2025},
  volume    = {73},
  number    = {10},
  pages     = {9886--9902},
  doi       = {10.1109/TCOMM.2025.3581967}
}

@misc{gong2024,
  author       = {Gong, Zijun and Jiang, Fan and Song, Yuhui and Li, Cheng and Tao, Xiaofeng},
  title        = {{Channel Estimation, Interpolation and Extrapolation in Doubly-Dispersive Channels}},
  howpublished = {arXiv:2408.09381 [eess.SP]},
  year         = {2024}
}

@ARTICLE{Wu2023,
  author    = {Wu, Yongzhi and Han, Chong and Chen, Zhi},
  journal   = {IEEE Trans. Wireless Commun.},
  title     = {{{DFT}-Spread Orthogonal Time Frequency Space System With Superimposed Pilots for Terahertz Integrated Sensing and Communication}},
  year      = {2023},
  volume    = {22},
  number    = {11},
  pages     = {7361--7376},
  doi       = {10.1109/TWC.2023.3250267}
}

@misc{doha2025,
  author       = {Doha, Shadman Rahman and Abdelhadi, Ahmed},
  title        = {{Deep Learning in Wireless Communication Receiver: A Survey}},
  howpublished = {arXiv:2501.17184 [cs.IT]},
  year         = {2025}
}

@ARTICLE{Albreem2021,
  author    = {Albreem, Mahmoud A. and Salah, Wael and Kumar, Arun and Alsharif, Mohammed H. and Rambe, Ali Hanafiah and Jusoh, Muzammil and Uwaechia, Anthony Ngozichukwuka},
  journal   = {IEEE Access},
  title     = {{Low Complexity Linear Detectors for Massive {MIMO}: A Comparative Study}},
  year      = {2021},
  volume    = {9},
  pages     = {45740--45753},
  doi       = {10.1109/ACCESS.2021.3065923}
}

\end{document}